\documentclass[a4paper,USenglish,cleveref,autoref,thm-restate]{lipics-v2021}
\hideLIPIcs\nolinenumbers

\usepackage{color}
\usepackage{cite}

\usepackage{algorithm}
\usepackage[noEnd=true,indLines=false,commentColor=blue]{algpseudocodex}
\usepackage{graphicx}
\usepackage{subcaption}
\usepackage{makecell} 
\usepackage[normalem]{ulem}

\algnewcommand\inputm{{\sc input}} 
\algnewcommand\branch{{\sc branch}}
\algnewcommand\echo{{\sc echo}}
\algnewcommand\echob{{\sc echo2}} 
\algnewcommand\echoc{{\sc echo3}}
\algnewcommand\echod{{\sc echo4}}
\algnewcommand\echoe{{\sc echo5}}
\algnewcommand\report{{\sc report}}

\title{Multi-Valued Connected Consensus:\\
A New Perspective on Crusader Agreement\\and Adopt-Commit}
\titlerunning{Multi-Valued Connected Consensus}
\author{Hagit Attiya}{Department of Computer Science, Technion, Israel}{}{}{}
\author{Jennifer L.\ Welch}{Department of Computer Science and Engineering,
   Texas A\&M University, USA}{}{}{}
\authorrunning{Attiya and Welch}
\ccsdesc{Theory of computation~Distributed algorithms}
\keywords{graded broadcast, gradecast, 
binding, approximate agreement}
\begin{document}

\maketitle

\begin{abstract}
Algorithms to solve fault-tolerant consensus in asynchronous systems
often rely on primitives such as crusader agreement, adopt-commit, and
graded broadcast, which provide weaker agreement properties than consensus.
Although these primitives have a similar flavor, they have been defined
and implemented separately in ad hoc ways.
We propose a new problem called \emph{connected consensus} that has as
special cases crusader agreement, adopt-commit, and graded broadcast,
and generalizes them to handle multi-valued inputs.
The generalization is accomplished by
relating the problem to approximate agreement on graphs.

We present three algorithms for multi-valued connected consensus in
asynchronous message-passing systems, one tolerating crash failures
and two tolerating malicious (unauthenticated Byzantine) failures.  We
extend the definition of {\it binding}, a desirable property recently
identified as supporting binary consensus algorithms that are correct
against adaptive adversaries, to the multi-valued input case and show
that all our algorithms satisfy the property.  Our crash-resilient
algorithm has failure-resilience and time complexity that we show are
optimal.  When restricted to the case of binary inputs, the algorithm
has improved time complexity over prior algorithms.  Our two
algorithms for malicious failures trade off failure resilience and
time complexity.  The first algorithm has time complexity that we
prove is optimal but worse failure-resilience, while the second has
failure-resilience that we prove is optimal but worse time complexity.
When restricted to the case of binary inputs, the time complexity (as
well as resilience) of the second algorithm matches that of prior
algorithms. 
\end{abstract}

\section{Introduction}

One way to address the impossibility of solving consensus in asynchronous
systems~\cite{FischerLP1985}
is to employ unreliable \emph{failure detectors}~\cite{ChandraT1996}.
Several algorithms in this class (e.g.,~\cite{Gafni1998,BouzidMR2015})
combine a failure detector with a
mechanism for detecting whether processes have reached unanimity,
in the form of an \emph{adopt-commit} protocol~\cite{YangNG1998structured}.
In such a protocol, each process starts with a binary input value
and returns a pair $(v,g)$ where $v$ is one of the input values and
$g$ is either 1 or 2.
The process is said to pick $v$ as its output value;
furthermore, if $g = 2$, then it \emph{commits} to $v$,
and if $g = 1$, then it \emph{adopts} $v$.
In addition to the standard validity property that the output value
is the input of some correct process, an adopt-commit protocol ensures
that processes commit to at most one value, and if any
process commits to a value, then no process adopts the other value.

Another way to address the impossibility of consensus is to use randomization
and provide only probabilistic termination.
Some algorithms in this class (e.g.,~\cite{Toueg1984})
rely on a mechanism called \emph{crusader agreement}~\cite{Dolev1982byzantine}:
Roughly, if all processes start with the same value $v$,
they must decide on this value, and otherwise,
they may pick an \emph{undecided} value, denoted $\bot$.
Other algorithms in this class (e.g.,~\cite{DeligiosHLZ2021})
rely on \emph{graded broadcast}~\cite{FeldmanM1988},
also called \emph{graded crusader agreement},
\emph{graded consensus}, or just \emph{gradecast}.
In a sense, graded broadcast is a combination of adopt-commit
with crusader agreement:
the decisions are either $(v,g)$, where $v$ is a binary value
and $g$ is either 1 or 2, or $\bot$
(also denoted $(\bot,0)$).
As in adopt-commit, the requirement is that processes {\em commit} to at
most one value, but in addition, if  any process \emph{adopts}
a value, then no process {\em adopts} the other value.
In a sense, the $\bot$ value allows a separation between adopting
one value and adopting a different value.

\begin{figure}
\includegraphics[scale=0.4]{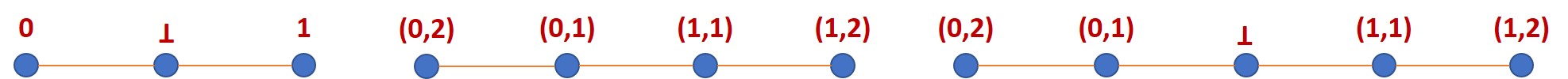}
\caption{Left:  crusader agreement.
Center:  adopt-commit.
Right:  graded broadcast.}
\label{fig:relate-tasks}
\end{figure}

Crusader agreement, adopt-commit and graded broadcast have a very
similar flavor, yet it is hard to tell them apart or
to pinpoint how they relate to each other.
(For example, some agreement protocols, e.g.,~\cite{BlumKLZL2020,MomoseR2022},
state that they use graded broadcast while, in fact,
they rely on an adopt-commit protocol, without a $\bot$ value.)
The relation between these primitives
becomes apparent when they are pictorially represented,
as in Figure~\ref{fig:relate-tasks},
with the possible decisions represented by vertexes on a path.
The different ``convergence'' requirements are all special cases of
the requirement that processes should decide on
the \emph{same or adjacent vertices} in the relevant graph.

With binary inputs, this description of the problems resembles
\emph{approximate agreement on the $[0,1]$ real interval with
parameter $\epsilon$}~\cite{DolevLPSW1986reaching}:
processes start at the two extreme points of the interval, 0 or 1,
and must decide on values that are at most $\epsilon$ apart from each other.
Decisions must also be valid, i.e., contained in the interval of the inputs.
Approximate agreement is a way to sidestep the impossibility
of solving consensus in asynchronous systems and
there are many algorithms for approximate agreement
(e.g.,~\cite{AbrahamAD2004optimal,Fekete1994asynchronous,Fekete1990asymptotically,DolevLPSW1986reaching}).

Indeed, crusader agreement reduces to approximate agreement with $\epsilon = \frac{1}{2}$:
Run approximate agreement with your input (0 or 1) to get some output $y$,
then choose the value in $\{0, \frac{1}{2}, 1\}$ that is closest to $y$
(taking the smaller one if there are two such values, e.g., for $y = \frac{1}{4})$.
Finally, return $\bot$ if $\frac{1}{2}$ is chosen.
(A similar observation is noted in~\cite{MahaneyS1985inexact,Fekete1990asymptotically}.)
Likewise,
adopt-commit reduces to approximate agreement with  $\epsilon = \frac{1}{3}$,
and graded consensus to taking $\epsilon = \frac{1}{4}$.
This connection makes it clear why \emph{binary}
crusader agreement, adopt-commit and graded broadcast can be
solved in an asynchronous message-passing system,
in the presence of crash and malicious (unauthenticated Byzantine) failures,
within a small number of communication rounds.

In some circumstances, agreement must be reached on a \emph{non-binary}
value, e.g., the identity of a leader~\cite{BenOrEY2003},
or the next message to deliver in totally-ordered atomic broadcast~\cite{CorreiaNV2006}.
This requires handling \emph{multi-valued} inputs,
where processes can start with an input from some set $V$
with $|V| \ge 2$.
We take inspiration from
\emph{approximate agreement on graphs}~\cite{CastanedaRR2018}, in which
each process starts with a vertex of a graph as its input and
must decide on a vertex such that all decisions are
\emph{within distance one of each other}
and
\emph{inside the convex hull of the inputs}.
When all processes start with the same node, this implies
they must decide on this node.

This paper defines a new problem, which we call \emph{connected consensus}.
Connected consensus elegantly unifies seemingly-diverse problems,
including crusader agreement, graded broadcast, and adopt-commit,
and generalizes them to accept \emph{multi-valued} inputs.
It can be viewed as approximate agreement on a restricted class of graphs.
Specifically, we focus on special cases of \emph{spider graphs}~\cite{Koebe1992}
consisting of a central clique (which could be a single vertex) to which
are attached $|V|$ paths (``branches'')
of length $R$, where $R$ is the {\em refinement} parameter.
(See Figures~\ref{fig:spider} and~\ref{fig:centerless spider}
in Section~\ref{sec:problem}.)

Recently, the definition of binary (graded) crusader agreement was
extended to include a \emph{binding} property~\cite{AbrahamBDY2022efficient}:
once the first correct process terminates, there exists a value
$v \in \{0,1\}$ such that no nonfaulty process
can output $v$ in any extension.
That paper demonstrates that this property facilitates
the modular design of
randomized consensus algorithms that
tolerate an \emph{adaptive} adversary.
We refer to~\cite{AbrahamBDY2022efficient} for an excellent description of
the usage, and its pitfalls, of (graded) crusader agreement,
together with a shared random coin, in randomized consensus;
they show how faster (graded) crusader agreement algorithms
lead to faster randomized consensus algorithms.
We generalize the binding property to hold for multi-valued
inputs:  once the first process decides, one value is ``locked'',
so that in all possible extensions, the decisions are on the same
branch of the spider graph.
Although the generalization is natural, it is worth pointing out that
simply applying the original definition unchanged when inputs are multi-valued
does not accomplish the desired goal when connected consensus is 
composed with a multi-valued shared random coin~\cite{CohenFGPZ2023}.
(See Section~\ref{sec:problem}.)

With these definitions at hand, we turn to designing algorithms for
multi-valued connected consensus in asynchronous message-passing systems
that tolerate crash or malicious failures and satisfy binding.
There is an algorithm for approximate agreement on general graphs
in the presence of malicious failures~\cite{NowakR2019}.
However, it requires exponential local computation
and does not satisfy the binding property.
We are interested in special-case spider graphs, as
described above; furthermore, we focus on the cases when the refinement
parameter $R$ equals either 1 or 2, which captures the applications of
interest.
Thus we exploit opportunities for optimizations to obtain better algorithms.

We present an algorithm for $R = 1$ and $R = 2$ with the binding
property that tolerates crash failures; it requires $n > 2f$,
where $n$ is the total number of
processes and $f$ is the maximum number of faulty processes.
(Appendix~\ref{sec:resilience lb} contains a simple proof
that $n > 2f$ is necessary in this case.)
Its time complexity is $R$ and its total message complexity is $O(n^2)$.
(As shown in Appendix~\ref{sec:time lb}, the time complexity is
optimal for reasonable resiliencies.)
The best previous algorithms, in \cite{AbrahamBDY2022efficient},
have slightly worse time complexity:  2 for $R = 1$ (crusader agreement) and 3
for $R = 2$ (graded crusader agreement).
Furthermore, both of these previous algorithms are for the binary case
($|V| = 2$) only, and cannot easily handle multiple values.

For malicious failures, we first present a simple algorithm with binding
for $R = 1$ and $R = 2$, that assumes $n > 5f$.
Like the crash-tolerant algorithm, its time complexity is $R$
and its total message complexity is $O(n^2)$.
(As shown in Appendix~\ref{sec:time lb}, the time complexity is
optimal for reasonable resiliencies.)
Both this algorithm and the crash-tolerant algorithm derive the binding
property from the inputs of the processes.
That is, the assignment of input values to the processes
uniquely determines which non-$\bot$ value, if any, can be decided in
any execution with that input assignment.
The fact that the locked value for binding is determined solely by the inputs
is conducive to the development of simple and efficient algorithms.
However, we show
that in the presence of malicious failures, the locked value cannot be
determined solely by the inputs
when $n < 5f$,
even if faulty processes do not equivocate (see Appendix~\ref{subsec:lb-rb}).

Our main algorithmic contribution
is a connected consensus algorithm for $R = 1$ and $R = 2$
with binding that tolerates malicious failures.
Its failure resilience is $n >3f$;
a simple proof (Appendix~\ref{sec:resilience lb})
shows that this is the optimal resilience.
Its time complexity is 5 for $R = 1$ and 7 for $R =2$, and its
total message complexity is $O(|V| \cdot n^2)$, where $V$ is the set of
input values.
The message complexity can be reduced to $O(n^2)$, at the cost of
increasing the time complexity by 2,
using techniques of~\cite{MostefaouiR2017}.

Figure~\ref{fig:summary} compares our algorithms with prior work.
The best previous algorithms with optimal resilience
are in~\cite{AbrahamBDY2022efficient} and are for the binary case only.
In~\cite{AbrahamBDY2022efficient}, the algorithms
are evaluated in terms of
``rounds'', giving smaller numbers than our time complexity measure;
we discuss the relationship between the two measures
at the end of Section~\ref{sec:malicious 5f}.

\begin{figure}
\centering
\begin{small}
\begin{tabular}{|l||c|c||c|c|c|c|}
\hline
failure type
  & \multicolumn{2}{c||}{crash}
  & \multicolumn{4}{c|}{malicious} \\
\hline
algorithm
  & Alg.~\ref{alg:crash}
  & \makecell{\cite[Algs.\ 3 \& 5]{AbrahamBDY2022efficient} \\ ($|V| = 2$)}
  & Alg.~\ref{alg:byz-5f}
  & Alg.~\ref{alg:byz-3f-binding}
  & Alg.~\ref{alg:byz-3f-binding} + \cite{MostefaouiR2017}
  & \makecell{\cite[Algs.\ 4 \& 6]{AbrahamBDY2022efficient} \\ ($|V| = 2$)} \\
\hline
\hline
resilience
  & $n > 2f$
  & $n > 2f$
  & $n > 5f$
  & $n > 3f$
  & $n > 3f$
  & $n > 3f$ \\
\hline
messages
  & $O(n^2)$
  & $O(n^2)$
  & $O(n^2)$
  & $O(|V|\cdot n^2)$
  & $O(n^2)$
  & $O(n^2)$ \\
\hline
time ($R = 1$) 
  & 1
  & 2
  & 1
  & 5
  & 7
  & 5 \\
\hline
time ($R = 2$) 
  & 2
  & 3
  & 2
  & 7
  & 9
  & 7 \\
\hline
\end{tabular}
\end{small}
\caption{Summary of connected consensus algorithms for $R = 1$
  (crusader agreement) and $R = 2$ (graded broadcast) with input set
  $V$; $n$ is the total number of processes, $f$ is the maximum number of
  faulty processes.  All algorithms satisfy Binding.}
\label{fig:summary}
\end{figure}

To summarize, our contributions are the following:
\begin{itemize}

\item We define the \emph{connected consensus} problem,
with a numeric \emph{refinement} parameter $R$.
The problem can be reduced to real-valued approximate agreement in the
binary case, and is equal to approximate agreement on
a specific class of \emph{spider} graphs in the multi-valued case.

\item We define the \emph{binding property} for the multi-valued
case, which previously was only defined for the binary case.

\item These insights lead us to design
efficient message-passing algorithms for connected consensus
with $R = 1$ or $2$,
in the presence of crash and malicious failures, for arbitrarily large
input sets.
The algorithms are modular in that the $R = 2$ case is obtained
by appending more communication exchanges to the $R = 1$ case.

\item For crash failures, our simple algorithm is optimal in resilience,
time complexity (for reasonable resiliencies), and message complexity.
Its time complexity improves on the best
previously known algorithms, which only handle binary inputs.

\item For malicious failures, we provide two algorithms that trade off
resilience and time and message complexity.
One algorithm has time complexity 1 or 2 (for $R = 1$ or $R = 2$),
which is optimal for reasonable resiliencies,
and sends $O(n^2)$ messages, but requires $n > 5f$.
The other algorithm only requires $n > 3f$, but has time complexity
5 or 7 (for $R = 1$ or $R = 2$) and sends $O(|V| \cdot n^2)$ messages.
This is the same performance as the algorithms
in~\cite{AbrahamBDY2022efficient} which are only for the case when $|V| = 2$.
\end{itemize}

\section{Model of Computation}
\label{sec:model}

We assume the standard asynchronous model for $n$ processes, up to
$f$ of which can be faulty,
in which processes communicate via reliable point-to-point messages.
We consider two possible types of failures: \emph{crash} failures, when a
faulty process simply ceases taking steps, and \emph{malicious} failures, when
a faulty process can change state arbitrarily and send messages with
arbitrary content.

In more detail, we assume a set of $n$ processes, each modeled as a
state machine.
Each process has a subset of initial states, with one state corresponding to
each element of $V$, denoting its input.
The transitions of the state machine are triggered by events.
There are two kinds of {\em events}:  spontaneous wakeup and receipt of a
message.
A transition takes the current state of the process and incoming message
and produces a new state of the process and a set of messages to be
sent to any subset of the processes.
The state set of a process contains
a collection of disjoint subsets,
each one modeling the fact that a particular decision has been taken;
once a process enters the subset of states for a specific decision,
the transition function ensures that it never leaves that subset.

A \emph{configuration} of the system is a vector of process states,
one for each process, and a set of in-transit messages.
In an initial configuration, each process is in an initial state and no
messages are in transit.
Given a subset of at most $f$ processes that are ``faulty'' with the rest
being ``correct'', we define an {\em execution} as a sequence of alternating
configurations and events $C_0, e_1, C_1,\ldots$ such that:
\begin{itemize}
\item $C_0$ is an initial configuration.
\item The first event for each process is wakeup.  A correct process
      experiences exactly one wakeup, a crash-faulty process experiences
      at most one wakeup, and a malicious-faulty process can experience
      any number of wakeups.
\item Suppose $e_i$ is a step by (correct)\footnote{Here and throughout,
      the restriction to correct processes is only for the case of
      malicious failures.} process $p$
  and let $s$ and $M$ be the state and set of messages resulting from $p$'s
  transition function applied to $p$'s state in $C_i$ and $m$, if $e_i$ is
  the receipt of message $m$ (or nothing if $e_i$ is a wakeup event).
  Then the only differences between $C_i$ and $C_{i+1}$ are that $m$
  is no longer in transit, $M$ is in transit, and $p$'s state in $C_{i+1}$
  is $s$.
  If $p$ is malicious, then $s$ and $M$ can be anything.
\item Every message sent by a process to a correct process is eventually
  received and the receipt occurs after the recipient wakes up.
\end{itemize}

Since we consider all executions that satisfy the above properties, we
are capturing an ``adaptive adversary'', which can control the inputs,
choice of faulty processes, ordering of process steps, behavior of
malicious processes, and the message delays, depending on anything
that has happened so far in the execution.

We are interested in worst-case complexity measures.  For communication
complexity, we count the maximum, over all executions, of the number of
messages sent by all the (correct) processes.

We adopt the definition in~\cite{AttiyaW1998}
for time complexity in an asynchronous message-passing system.
We start by defining a timed
execution as an execution in which nondecreasing nonnegative integers
(``times'') are assigned to the events, with no two events by the same
process having the same time.  For each timed execution, we
consider the prefix ending when the last correct process decides, and
then scale the times so that the \emph{maximum time} that elapses between the
sending and receipt of any message between correct processes
is 1.  We define the time complexity as the maximum, over all such scaled timed
execution prefixes, of the time assigned to the last event
minus the latest time when any (correct) process wakes up.
For simplicity, we assume that the first wakeup event of each process
occurs at time 0.
This definition of time complexity is analogous to that
in~\cite{MostefaouiMR2015,MostefaouiR2017}, which measures the length of
the longest sequence of causally related messages.

\section{Definitions of Connected Consensus and Related Problems}
\label{sec:problem}

\subsection{Connected Consensus}

\begin{figure}
\includegraphics[scale=0.45]{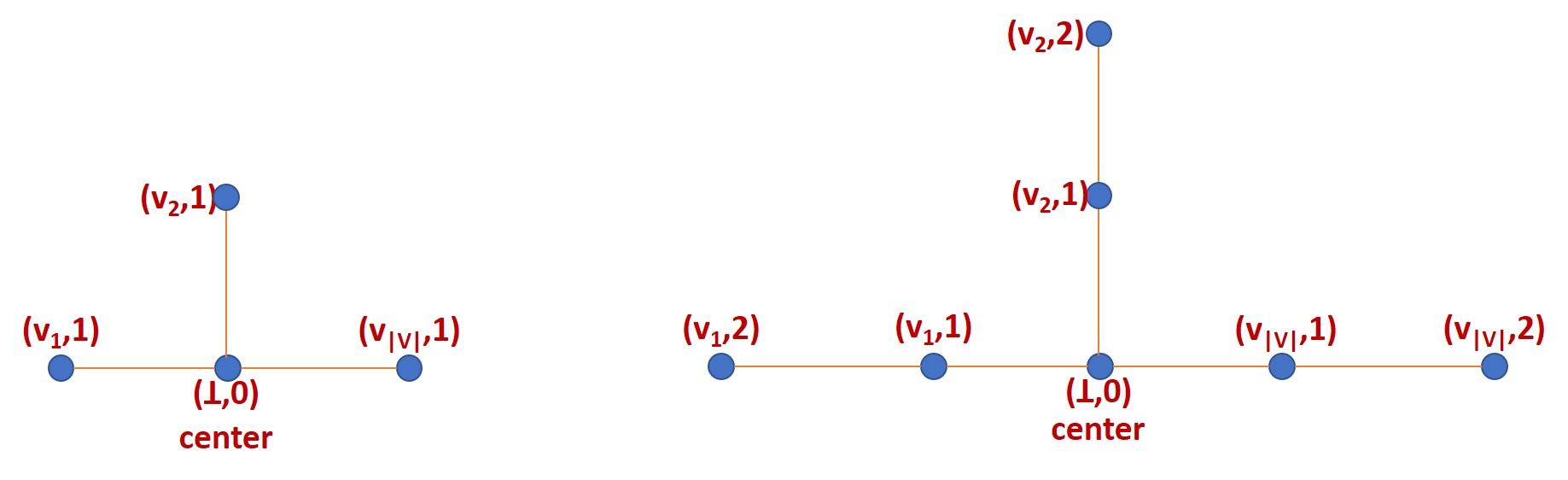}
\caption{Spider graphs: $R = 1$ (left) and $R = 2$ (right).}
\label{fig:spider}
\end{figure}

Let $V$ be a finite, totally-ordered set of values;
assume $\bot \notin V$.
Given a positive integer $R$, let $G_S(V,R)$ be the ``spider'' graph consisting
of a central vertex labeled $(\bot,0)$ that has $|V|$ paths extending from it,
with one path (``branch'') associated with each $v \in V$.
The path for each $v$ has $R$ vertices on it, not counting $(\bot,0)$, labeled
$(v,1)$ through $(v,R)$, with $(v,R)$ being the leaf.
(See Figure~\ref{fig:spider}.)
Given a subset $V'$ of $V$,
we denote by $T(V,R,V')$ the minimal subtree of $G_S(V,R)$ that connects the
set of leaves $\{(v,R) | v \in V'\}$;
note that when $V'$ is a singleton set $\{v\}$ then $T(V,R,\{v\})$
is the single (leaf) vertex $(v,R)$. 

In the \emph{connected consensus problem for $V$ and $R$}, each process
has an input from $V$.  The requirements are:

\begin{description}

\item{\bf Termination:}  Each correct process must decide on
a vertex of $G_S(V,R)$, namely, an element
of $\{(v,r) | v \in V, 1 \le r \le R\} \cup \{ (\bot,0) \}$.

\item{\bf Validity:} Let $I = \{(v,R) | v$ is the input of a
(correct)
process$\}$.
The output of each (correct) process must be a vertex in $T(V,R,I)$.
In particular, if all (correct) processes start with the same input $v$,
then $(v,R)$ must be decided.

\item{\bf Agreement:} The distance between the vertices labeled by the
decisions of all (correct) processes is at most one.
\end{description}

If we set $R = 1$, 
we get \emph{crusader agreement}~\cite{Dolev1982byzantine},
originally considered in the synchronous model.
If we set $R = 2$ 
we get \emph{graded broadcast}~\cite{FeldmanM1997},
originally considered in the synchronous model.
In asynchronous shared-memory systems, graded broadcast is also called
\emph{adopt-commit-abort}~\cite{DelporteFR2021,MostefaouiRRT2008}.

\subsection{Binding}

An additional condition of interest for the connected consensus
problem is called \emph{binding}~\cite{AbrahamBDY2022efficient}.
It was originally proposed for the case of binary inputs,
and defined as follows: 
``before the first non-faulty party terminates, there is a value $v \in
\{0,1\}$ such that no non-faulty party can output the value $v$ in any
continuation of the execution.''
Here we generalize this property for multi-valued inputs.

\begin{description}
\item{\bf Binding:}  In every execution prefix that ends with the
first (correct) process deciding, one value is ``locked'', meaning that
in every extension of the execution prefix, the decision of every
(correct) process must be on the same branch of the spider graph.
\end{description}

If the first decision is not $(\bot,0)$, then this condition
follows from Agreement.  More interestingly, if the first decision is
$(\bot,0)$, then there are many choices as to which branch is locked but
the choice must be the same in every extension.
Note that when $|V|=2$, our definition is equivalent to the original
from~\cite{AbrahamBDY2022efficient},
but for larger $V$, our definition is stronger---the original
definition only excludes one value, leaving $|V|-1$ possible
decision values,
while ours excludes $|V|-1$ values, leaving only one possible
decision value.

One example showing the original definition of binding is not
strong enough with multi-valued inputs is the framework for
solving Byzantine Agreement in an asynchronous system by alternating
calls to a (black box) connected consensus with $R=1$ (crusader agreement)
subroutine with calls to a
shared random coin subroutine (cf.~\cite[Section 3]{AbrahamBDY2022efficient}
and~\cite[Section 2]{AttiyaEN2023}).
Suppose there are three processes with inputs 0, 1 and 2,
and the first process, say $p$, return from connected consensus
gets $(\bot,0)$ (corresponding to $\bot$ in crusader agreement).
According to the original definition, some value $v \in \{0,1,2\}$
is no longer a possible output of the connected consensus subroutine.
Since $p$ obtains $(\bot,0)$ from crusader agreement, 
it calls the shared coin; 
let $v' \in \{0,1,2\}$ be the value $p$ get from the shared coin. 
However, the original binding property still allows the adversary 
to make the other processes return $(v'',1)$, with $v'' \neq v'$.
This would mean that processes start the next iteration in disagreement.

When $R = 1$, there are only two vertexes on any given branch of
the spider graph, $(v,1)$ and $(\bot,0)$.  This implies:

\begin{proposition}
\label{prop:binding-agreement}
If $R = 1$, then the Binding property implies the Agreement property.
\end{proposition}

If $R = 2$, though, the Binding property only restricts the branch of
the spider graph on which decisions can be made; both $(\bot,0)$ and
$(v,2)$ are on the same branch, but Agreement does not permit them to both
be decided.

\subsection{Centerless Variants}
\label{subsec:centerless}

\begin{figure}
\includegraphics[scale=0.4]{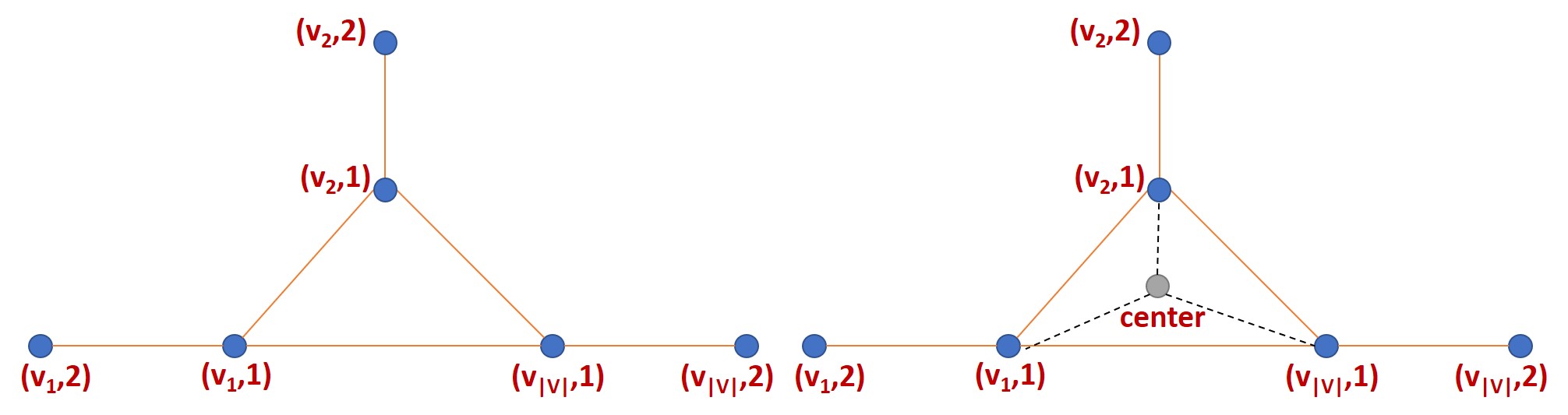}
\caption{Centerless spider graph with $R = 2$ (left)
and its reduction to a (centered) graph (right).}
\label{fig:centerless spider}
\end{figure}

Recall that in the \emph{adopt-commit}
problem~\cite{Gafni1998,YangNG1998structured},
processes return a pair $(v,g)$ where $v$ is one of the input values and
$g$ is either 1 (adopt) or 2 (commit).
Thus, there is no analog of the ``center'' vertex.
We model this with a \emph{centerless} spider graph
(see left side of Figure~\ref{fig:centerless spider}).
Here, $G_S(V,R)$ is the graph consisting of a clique on the vertices
$(v,1)$ for all $v \in V$, each with a path extending from it,
with $R-1$ vertices on it, not counting $(\bot,0)$, labeled
$(v,2)$ through $(v,R)$, with $(v,R)$ being the leaf.
Decisions must satisfy termination, validity and agreement as specified
for the variant with a center.
Since the graph has no center, binding cannot be defined;
indeed, when a process returns $(v,1)$,
other processes might return $(v',1)$, for $v \neq v'$.

Instead of developing algorithms directly for the centerless
problem, we note that it can be reduced to the centered problem
with the same refinement parameter:
Call the algorithm for the centered problem with your input $u$.
If the return value is $(v,g)$ with $g > 0$, then
decide this value for the centerless problem;
when the return value is $(\bot,0)$ \emph{(i.e, the center)},
decide $(u,1)$ for the centerless problem.
(See right side of Figure~\ref{fig:centerless spider}).
This reduction implies the following proposition:

\begin{proposition}
If $A$ is an algorithm that solves the (centered)
connected consensus problem for $R$, then there is an algorithm $A'$
that solves the centerless connected consensus problem for $R$.
\end{proposition}

In the \emph{vacillate-adopt-commit} (VAC) problem~\cite{AfekACV2017brief},
the possible output values are ($v$,commit), ($v$,adopt), and ($v$,vacillate),
where $v$ is any value.
If any output is ($v$,commit), then every other output is either
($v$,commit) or ($v$,adopt), for the same $v$.
Furthermore, if there is no commit output and there is at least one
($v$,adopt) output, then every other output is either ($v$,adopt),
with the same value $v$, or ($w$,vacillate), where $w$ can be any value.
At first glance, VAC seems to correspond to a centerless spider graph
with refinement parameter $R = 3$.
However, a closer look at the usage of VAC suggests
that 
the return value of vacillate is irrelevant,
in which case the problem could be
represented as a centered spider graph with $R = 2$,
like adopt-commit-abort and graded crusader agreement.

\section{Crash-Tolerant Algorithm}

We first note that a standard partitioning argument shows that $n >
2f$ is required to solve connected consensus with crash failures for
any $R \ge 1$, even without binding; see Appendix~\ref{sec:resilience lb}.

We next present an algorithm for connected consensus
with binding for $n$ processes that tolerates $f < n/2$ crash failures.
The algorithm is extremely simple and efficient, using one round of
message exchanges for $R = 1$ and two rounds of message exchanges
for $R = 2$, and thus using only $O(n^2)$ messages.
Due to the simple communication structure, the number of rounds
of message exchanges is equal to the time complexity.
In Appendix~\ref{sec:time lb}, we use a reduction from
approximate agreement\footnote{To solve $\epsilon$-approximate
  agreement, we use a subroutine for connected consensus with $V =
  \{0,1\}$ and $R = \left\lceil \frac{1}{2\epsilon} \right\rceil$.
  The approximate agreement input, which is either 0 or
  1, is used as the input to the connected consensus subroutine.  To
  obtain the approximate agreement output, map the $2R+1$ vertices of the
  connected consensus graph, which is a chain, in order to points in the real
  interval $[0,1]$ that are equally spaced, with $(0,R)$ corresponding
  to 0 and $(1,R)$ to 1.  Since adjacent points in $[0,1]$ are $\frac{1}{2R}$
  apart, they are within $\epsilon$ of each other.}
to derive that the time complexity
for $R = 1$ is optimal and show that for $R = 2$, it is optimal as long
$n \le 4f$.  Interestingly, if $n > 4f$, there is a one-round algorithm
for $R = 2$.

In the first round, processes exchange their inputs.
After hearing from $n-f$ processes,
each process chooses branch $v$ of the spider graph, if all the received
values equal $v$, or the center vertex otherwise.
If $R = 1$, then the process decides on the branch.
Otherwise, processes exchange branch values ($v$ or $\bot$)
in a second round in order to decide on a vertex on the $v$ branch.
After waiting for $n-f$ messages in the second round,
if a process' branch is $\bot$, then it decides $(v,1)$ if at least
one second-round message is for $v$ and $(\bot,0)$ otherwise.
If the process' branch is $v$, then it decides $(v,2)$ if all the
second-round messages are for $v$ and $(v,1)$ otherwise.
The pseudocode is in Algorithm~\ref{alg:crash}.

\begin{algorithm}[tb]
\caption{Connected Consensus algorithm with Binding for $R = 1,2$ with
   $n$ processes, $f < n/2$ of which may crash; code for process $p$}
\label{alg:crash}
\begin{algorithmic}[1]
\State send $\langle$\inputm,input$\rangle$ to all \Comment{round 1}
\State wait for $n-f$ \inputm{} messages
\State let $W$ be set of values received in \inputm{} messages
\If{$\exists v \in V$ s.t. $W = \{v\}$}
   \State branch $:= v$
\Else
   \State branch $:= \bot$
\EndIf
\If{$R = 1$}
   \If{branch $=\bot$}
      \State decide $(\bot,0)$ \Comment{center vertex}
   \Else
      \State decide (branch,1) \Comment{leaf vertex for chosen branch}
   \EndIf
\Else \Comment{$R = 2$; round 2}
   \State send $\langle$\branch,branch$\rangle$ to all
   \State wait for $n-f$ \branch{} messages
   \If{branch $=\bot$}
      \If{$\exists v \in V$ s.t. at least one \branch{} message has value $v$}
         \State decide $(v,1)$ \Comment{middle vertex on branch for $v$}
                               \label{line:alg-crash:dec-1a}
      \Else
         \State decide $(\bot,0)$ \Comment{center vertex}
      \EndIf
   \Else \Comment{branch $\ne \bot$}
      \If{$\exists v \in V$ s.t. all \branch{} messages have value $v$}
         \State decide $(v,2)$ \Comment{leaf vertex for $v$}
      \Else
         \State decide (branch,1) \Comment{middle vertex on branch chosen
                                           in round 1}\label{line:alg-crash:dec-1b}
      \EndIf
   \EndIf
\EndIf
\end{algorithmic}
\end{algorithm}

\begin{restatable}{theorem}{crashtwo}
\label{thm:alg-crash-correct}
If $n > 2f$, then Algorithm~\ref{alg:crash} solves binding connected
consensus for $R = 1$ and $R = 2$
with $n$ processes, up to $f$ of which can crash.
It takes 1 time unit and sends $O(n^2)$ messages for $R = 1$
and takes 2 time units and sends $O(n^2)$ messages for $R = 2$.
\end{restatable}

The proof of this theorem appears in Appendix~\ref{sec:crash proof}.
Here, we only outline why the algorithm is binding.
In fact, we show a stronger property, that the branch along which
decisions are made is determined solely by the inputs.

For any assignment of inputs to the processes,
since $n > 2f$,
there is at most one input value $v \in V$ that occurs at least
$n-f$ times.
Note that if $p$ sets its branch variable to $v$,
then all $n-f$ \inputm{} messages it receives are for $v$,
and since processes fail only by crashing,
the $n-f$ senders of these messages all have input $v$.
Therefore, no process can set its branch variable to
any value in $V$ other than $v$.
For $R = 1$, processes decide on their branch variables,
implying that in any future extension, every process that sets its
branch variable sets it to $v$ or $\bot$.
For $R = 2$, processes exchange their branch variables in the second
round.  The only possible non-$\bot$ value that can be exchanged is $v$,
so the only possible decisions are $(\bot,0)$, $(v,1)$ and $(v,2)$,
which proves binding for $R = 2$.

\section{Algorithms Tolerating Malicious Failures}

We first note that $n > 3f$ is required to solve connected consensus
for any $R \ge 1$ with malicious failures, even without binding;
a proof appears in Appendix~\ref{sec:resilience lb}.

We next present two algorithms for connected consensus,
tolerating malicious failures.
We start with a simple algorithm,
for $n > 5f$ using ideas from~\cite{DolevLPSW1986reaching}.
Then we present a more complex algorithm for $n > 3f$,
which is a modular extension of algorithms in~\cite{AbrahamBDY2022efficient}
incorporating new ideas inspired by~\cite{MostefaouiR2017}
to deal with multi-valued inputs.
Although the $n > 3f$ algorithm has better resilience than the $n > 5f$
algorithm, it has worse time complexity.
In Appendix~\ref{sec:time lb}, we use the same reduction from
approximate agreement as we did for crash failures
to derive  that the time complexity
for $R = 1$ is optimal and show that for $R = 2$, it is optimal as long
as $n \le 9f$.  For higher resiliencies, in particular if
$n > 13f$, there is a one-round algorithm with binding for $R = 2$.

\subsection{Algorithm for $n > 5f$}

In this subsection we present an algorithm for connected consensus
with binding for $n$ processes that tolerates $f < n/5$ malicious failures.
The algorithm is extremely simple and efficient, using one round of
message exchanges for $R = 1$ and two rounds of message exchanges
for $R = 2$, and thus using only $O(n^2)$ messages.

In the first round, processes exchange their inputs and, after
hearing from $n-f$ processes, each process drops the $f$ smallest and $f$
largest values received, an idea inspired by approximate agreement algorithms
(e.g.,~\cite{DolevLPSW1986reaching}).
Then each process chooses branch $v$ of the spider graph, if all the remaining
values equal $v$, or the center vertex otherwise.
If $R = 1$, then the process decides on the branch.
Otherwise, processes exchange branch values ($v$ or $\bot$)
in a second round in order to decide on a vertex on the $v$ branch.
This is done in a manner that is similar to the second round
in our crash-resilient algorithm, Algorithm~\ref{alg:crash}.
After waiting for $n-f$ messages in the second round,
if a process' branch is $\bot$, then it decides $(v,1)$ if at least
$f+1$ second-round messages are for $v$ and $(\bot,0)$ otherwise.
If the process' branch is $v$, then it decides $(v,2)$ if at least $n-2f$
second-round messages are for $v$ and $(v,1)$ otherwise.
The pseudocode, which is similar to Algorithm~\ref{alg:crash},
appears in Algorithm~\ref{alg:byz-5f}.

\begin{algorithm}[tb]
\caption{Connected Consensus algorithm with Binding for $R = 1, 2$
  with $n$ processes, $f < n/5$ of which may be malicious;
  code for process $p$}
\label{alg:byz-5f}
\begin{algorithmic}[1]
\State send $\langle$\inputm,input$\rangle$ to all \Comment{round 1}
\State wait for $n-f$ \inputm{} messages
\State let $W$ be multiset of values received in \inputm{} messages,
       dropping $f$ smallest and $f$ largest
\If{$\exists v \in V$ s.t. every element in $W$ is $v$}
   \State branch $:= v$
\Else
   \State branch $:= \bot$
\EndIf
\If{$R = 1$}
   \If{branch $= \bot$}
      \State decide $(\bot,0)$  \Comment{center vertex}
   \Else
      \State decide (branch,1)  \Comment{leaf vertex for chosen branch}
   \EndIf
\Else \Comment{$R = 2$; round 2}
   \State send $\langle$\branch,branch$\rangle$ to all
   \State wait for $n-f$ \branch{} messages
   \If{branch $= \bot$}
      \If{$\exists v \in V$ s.t. at least $f+1$ \branch{} messages
             have value $v$}
         \State decide $(v,1)$ \Comment{middle vertex on branch for $v$}
                 \label{line:alg-5f:dec-1a}
      \Else
         \State decide $(\bot,0)$ \Comment{center vertex}
      \EndIf
   \Else \Comment{branch $\ne \bot$}
      \If{$\exists v \in V$ s.t. at least $n-2f$ \branch{} messages
              have value $v$}
         \State decide $(v,2)$ \Comment{leaf vertex for $v$}
      \Else
         \State decide (branch,1)
                 \Comment{middle vertex on branch chosen in round 1}
                 \label{line:alg-5f:dec-1b}
      \EndIf
   \EndIf
\EndIf
\end{algorithmic}
\end{algorithm}

\begin{restatable}{theorem}{malfive}
\label{thm:alg-5f-correct}
If $n > 5f$, then Algorithm~\ref{alg:byz-5f} solves binding connected
consensus for $R = 1$ and $R = 2$
with $n$ processes, up to $f$ of which can be malicious.
It takes 1 time unit and sends $O(n^2)$ messages for $R = 1$
and takes 2 time units and sends $O(n^2)$ messages for $R = 2$.
\end{restatable}

The proof of this theorem appears in Appendix~\ref{sec:5f proof}.
Here, we only outline why the algorithm is binding,
which similarly to Algorithm~\ref{alg:crash} follows from a
stronger property, that the branch along which
decisions are made is determined solely by the inputs.

Given an assignment of inputs to the processes,
suppose there is one execution in which a correct process decides $u \in V$
and another execution in which a correct process decides $v \in V$
with $u < v$.
It can be shown (see Lemma~\ref{lem:inputs-branch-5f}(a)),
at least $n-3f$ correct processes have input at most $u$, and
that at least $n-3f$ correct processes have inputs that are at least $v$.
Thus the total number of processes $n$ is at least $2(n-3f)$ plus
the $f$ faulty processes.  That is, $n \ge 2(n-3f) + f$, which implies
$n \le 5f$, a contradiction.
This implies binding for $R=1$.
For $R=2$, this argument shows that there is only one possible $v \in V$
that can appear in correct
processes' branch variables at the end of round 1 in any execution.
Thus no correct process can get more than $f$ \branch{} messages for
any $u \in V$ other than $v$ and thus it cannot decide $(u,1)$ or $(u,2)$
in any execution.

\subsection{Algorithm for $n > 3f$}
\label{sec:malicious 5f}

In this subsection we present an algorithm for connected consensus
with binding for $n$ processes that tolerates $f < n/3$ malicious
failures.  The time complexity for $R = 1$ is 5 and for $R = 2$ is 7,
while the message complexity is $O(|V| \cdot n^2)$ in both cases.
The failure-resiliency is optimal, per the discussion
at the beginning of this section.
The $|V|$ factor in the message complexity can be reduced to a constant,
resulting in $O(n^2)$ message complexity,
by first employing the RD-broadcast primitive in~\cite{MostefaouiR2017}, which
reduces the number of values under consideration to 6, at the cost of
$O(n^2)$ additional messages and two additional time units.

The algorithm is a modular combination of Algorithms 4 (for $R = 1$) and 6
(for $R = 2$) in~\cite{AbrahamBDY2022efficient}, which work when $|V| = 2$,
with the addition of a mechanism from the
MV-broadcast in~\cite{MostefaouiR2017} to handle $|V| > 2$.
MV-broadcast is a primitive to reduce the number of input values under
consideration to two in the context of solving consensus.

Processes exchange input values in increasing ``levels''
of \echo{} messages.
Initially, processes exchange their inputs via \echo{} messages
and also use \echo{} messages to amplify values that have been received
at least $f+1$ times; this threshold ensures that at least one correct
process has that value as its input.
To handle the situation when $|V| > 2$ and there may not be any message
that is sent in at least $f+1$ \echo{} messages, an \echo{} message
for $\bot$ is initiated if a process receives at least $f+1$ \echo{}
messages in addition to those for the most commonly received value;
this condition only holds if there are at least two different input
values at the correct processes.
This early appearance of $\bot$ requires some later modifications,
mentioned below, to the original algorithm.

Whenever a process receives $n-f$ \echo{} messages for some value $v$,
it stores $v$ in its local set variable ``approved'';
the first time
this happens, it sends the value in an \echob{} message.
The $n-f$ threshold ensures some level of uniformity in processes'
``approved'' sets.
Each process sends one \echoc{} message, either for $\bot$ if it
collects more than one approved value, or for value $v$ if it
receives $n-f$ \echob{} messages for $v$.
The \echoc{} messages have the desirable property that only one
non-$\bot$ value is sent in them by the correct processes.
When $R = 1$, processes decide once at least $n-f$ \echoc{} messages
have been received: if there are at least $n-f$ for the same value
$v$, then $(v,1)$ is decided, otherwise if there are at least two
approved values or if $\bot$ is approved\footnote{Checking if $\bot$
has been approved here and later are modifications needed because of the
possibility that $\bot$ is sent in \echo{} messages.},
then $(\bot,0)$ is decided.

When $R = 2$, processes continue for two more levels in order to refine
the values obtained so far on the chosen branch.
The value chosen as the decision in the $R = 1$ case is sent in an \echod{}
message; these message inherit the property that at most one non-$\bot$
appears in those sent by correct processes.
Each process waits for \echod{} messages.
If eventually it collects $n-f$ for a common value, then it sends an
\echoe{} message for that value; if eventually it has more than one
approved value or if $\bot$ is approved,
it sends an \echoe{} message for $\bot$.
\echoe{} messages also inherit the property that at most one non-$\bot$
appears in those sent by correct processes.

The decision is based on \echod{} and \echoe{} messages received.
If a process receives at least $n-f$ \echoe{} messages for the same
non-$\bot$ value, then it decides $(v,2)$.
If it receives at least $n-f$ \echoe{} messages for $\bot$, then
it decides $(\bot,0)$.
Otherwise, if it has approved either $\bot$ or at least two values,
receives at least one \echoe{} message for some non-$\bot$
value $w$, and has at least $f+1$ \echod{} messages for $w$, it decides
$(w,1)$.

The pseudocode is in Algorithm~\ref{alg:byz-3f-binding}.
The presentation differs from that in our Algorithms~\ref{alg:crash}
and~\ref{alg:byz-5f} and in~\cite{AbrahamBDY2022efficient}.
Instead of using syntactic constructs such as ``wait until'' and ``upon''
receiving certain messages, our code is purely interrupt-driven in order
to clarify the interactions between the receipts of different messages
and the conditions triggering various actions.
The technique inspired by~\cite{MostefaouiR2017} appears in Lines
\ref{line:alg-3f-binding:start-new} through
\ref{line:alg-3f-binding:end-new}.
We denote by sum$(A)$, where $A$ is an array of integers,
the sum of all the entries in $A$.
A correct process sends at most one \echo{} message for any $v \in V \cup \{\bot\}$,
and at most one \echob, \echoc, \echod, and \echoe{} message.
This allows us to assume there is some mechanism for eliminating duplicate
messages that arrive from the same (faulty) sender.

\tikzset{algpxDefaultBox/.style={style}} 

\begin{algorithm}
\caption{Connected Consensus algorithm with Binding for $R = 1,2$ with
    $n$ processes, $f < n/3$ of which may be malicious; code for process $p$}
\label{alg:byz-3f-binding}
\begin{algorithmic}[1]
\State \underline{initially:}
\State approved $:= \emptyset$ (subset of $V \cup \{\bot\}$)
     \Comment{set of approved values}
\State num\_echo$[v] := 0$ for $v \in V \cup \{\bot\}$
     \Comment{\# received \echo{} messages for $v$}
\State num\_echo$i[v] := 0$ for $2 \le i \le 5$,
                                   $v \in V \cup \{\bot\}$
     \Comment{\# received \echo-$i$ messages for $v$}
\State sent\_echo$[v] :=$ false for $v \in V \cup \{\bot\}$
     \Comment{has $p$ sent an \echo{} message for $v$ yet?}
\State sent\_echo$i :=$ false for $2 \le i \le 5$
     \Comment{has $p$ sent an \echo-$i$ message yet?}
\begin{comment}
\State num\_echo$[v] := 0$ for all $v \in V \cup \{\bot\}$
     \Comment{\# received \echo{} messages for $v$}
\State num\_echo2$[v] := 0$ for all $v \in V \cup \{\bot\}$
     \Comment{\# received \echob{} messages for $v$}
\State num\_echo3$[v] := 0$ for all $v \in V \cup \{\bot\}$
     \Comment{\# received \echoc{} messages for $v$}
\State num\_echo4$[v] := 0$ for all $v \in V \cup \{\bot\}$
     \Comment{\# received \echod{} messages for $v$}
\State num\_echo5$[v] := 0$ for all $v \in V \cup \{\bot\}$
     \Comment{\# received \echoe{} messages for $v$}
\State sent\_echo$[v] :=$ false for all $v \in V \cup \{\bot\}$
     \Comment{has $p$ sent an \echo{} message for $v$ yet?}
\State sent\_echo2 $:=$ false
     \Comment{has $p$ sent an \echob{} message yet?}
\State sent\_echo3 $:=$ false
     \Comment{has $p$ sent an \echoc{} message yet?}
\State sent\_echo4 $:=$ false
     \Comment{has $p$ sent an \echod{} message yet?}
\State sent\_echo5 $:=$ false
     \Comment{has $p$ sent an \echoe{} message yet?}
\end{comment}
\State decided $:=$ false
     \Comment{has $p$ decided yet?}
\Statex  % ............................................................
\State \underline{wakeup:}
\State send $\langle$\echo,input$\rangle$ to all; sent\_echo[input] $:=$ true
     \Comment{initiate \echo{} for $p$'s input}
\Statex  % ............................................................
\State \underline{receive $\langle$\echo$,v\rangle$:}
\State num\_echo$[v]++$
\If{(num\_echo$[v] = f+1$) and (!sent\_echo$[v]$)}
    \State send $\langle$\echo$,v\rangle$ to all;
           sent\_echo$[v] :=$ true
           \Comment{echo $v$ if enough support but only once}
\ElsIf{(sum(num\_echo) $-$ num\_echo$[m] \ge f+1)$ and
       (!sent\_echo$[\bot]$),\\
             where $m$ is s.t.\ num\_echo$[m] \ge$ num\_echo$[u]$
             for all $u \in V \cup \{\bot\}$)}\\
    \Comment{if evidence for multiple correct inputs then initiate \echo{}
             for $\bot$} \label{line:alg-3f-binding:start-new}
    \State send $\langle$\echo$,\bot\rangle$ to all;
           sent\_echo$[\bot] :=$ true
           \label{line:alg-3f-binding:end-new}
\ElsIf{num\_echo$[v] = n-f$}
    \If{!sent\_echo2}   \Comment{send only one \echob}
        \State send $\langle$\echob$,v\rangle$ to all;
               sent\_echo2 $:=$ true
    \EndIf
    \State add $v$ to approved
    \If{($|$approved$| > 1$) and (!sent\_echo3)}
        \Comment{send only one \echoc}
        \State send $\langle$\echoc$,\bot\rangle$ to all;
               sent\_echo3 $:=$ true
    \EndIf
\EndIf
\Statex  % ............................................................
\State \underline{receive $\langle$\echob$,v\rangle$:}
\State num\_echo2$[v]++$
\If{(num\_echo2$[v] = n-f$) and (!sent\_echo3)}
        \Comment{send only one \echoc}
    \State send $\langle$\echoc$,v\rangle$ to all; sent\_echo3 $:=$ true
\EndIf
\Statex  % ............................................................
\State \underline{receive $\langle$\echoc$,v\rangle$:}
\State num\_echo3$[v]++$
\If{(sum(num\_echo3) $\ge n-f$) and
                (($|$approved$| > 1$) or ($\bot \in$ approved))}
    \If{($R=1$) and (!decided)}  \Comment{decide only once}
        \State decide $(\bot,0)$; decided $:=$ true  \Comment{center vertex}
    \ElsIf{($R=2$) and (!sent\_echo4)} \Comment{send only one \echod}
        \State send $\langle$\echod$,\bot \rangle$ to all;
               sent\_echo4 $:=$ true
    \EndIf
\ElsIf{num\_echo3$[v] \ge n-f$}
    \If{($R = 1$) and (!decided)} \Comment{decide only once}
        \State decide $(v,1)$; decided $:=$ true  \Comment{leaf vertex for $v$}
    \ElsIf{($R=2$) and (!sent\_echo4)} \Comment{send only one \echod}
        \State send $\langle$\echod,$v \rangle$ to all;
               sent\_echo4 $:=$ true
    \EndIf
\EndIf
\Statex \Comment{continued...................................}
\algstore{alg-3f-binding-split}
\end{algorithmic}
\end{algorithm}

\begin{algorithm}
\begin{algorithmic}
\algrestore{alg-3f-binding-split}
\Statex \Comment{...........................Continuation of Algorithm~\ref{alg:byz-3f-binding}}
\Statex
\State \underline{receive $\langle$\echod$,v\rangle$:}
\State num\_echo4$[v]++$
\If{(num\_echo4$[v] = n-f$) and (!sent\_echo5)}
                     \Comment{send only one \echoe}
    \State send $\langle$\echoe$,v\rangle$ to all;
           sent\_echo5 $:=$ true
\ElsIf{(sum(num\_echo4) $\ge n-f)$
                and (($|$approved$| > 1$) or ($\bot \in$ approved))
                and (!sent\_echo5)}
   \State send $\langle$\echoe$,\bot\rangle$ to all;
          sent\_echo5 $:=$ true
\EndIf
\Statex  % ............................................................
\State \underline{receive $\langle$\echoe$,v\rangle$:}
\State num\_echo5$[v]++$
\If{!decided}   \Comment{decide only once}
    \If{($v \in V$) and (num\_echo5$[v] \ge n-f$)}
        \State decide $(v,2)$; decided $:=$ true  \Comment{leaf vertex for $v$}
    \ElsIf{(sum(num\_echo5) $\ge n-f)$ and
                     (($|$approved$| > 1$) or ($\bot \in$ approved)) and \\
              there exists $w \in V$ s.t.\
              (num\_echo5$[w] \ge 1)$ and (num\_echo4$[w] \ge f+1)$}
              \label{line:alg-3f-binding:decide-intermediate}
        \State decide $(w,1)$; decided $:=$ true
                             \Comment{middle vertex on branch for $w$}
    \ElsIf{num\_echo5$[\bot] \ge n-f$}
        \State decide $(\bot,0)$   \Comment{center vertex}
    \EndIf
\EndIf
\end{algorithmic}
\end{algorithm}

\begin{restatable}{theorem}{malthree}
\label{thm:alg-3f-binding-correct}
If $n > 3f$, then Algorithm~\ref{alg:byz-3f-binding} solves binding
connected consensus for $R = 1$ and $R = 2$
with $n$ processes, up to $f$ of which can be malicious.
It takes 5 time units and sends $O(|V| \cdot n^2)$ messages for $R = 1$
and takes 7 time units and sends $O(|V| \cdot n^2)$ messages for $R = 2$.
\end{restatable}

The complete proof of Theorem~\ref{thm:alg-3f-binding-correct}
appears in Appendix~\ref{sec:3f proof} and is sketched below.

\begin{proof} (Sketch)
First, we argue that
the values sent in \echo{} messages by correct processes are ``valid'':
if the value is in $V$, then some correct process has input $v$,
whereas if the value is $\bot$, then not all the correct processes have
the same input.  The latter property is ensured by
lines~\ref{line:alg-3f-binding:start-new}
through~\ref{line:alg-3f-binding:end-new} for the following reason.
The first correct process to send \echo{} for $\bot$ does so because
it receives at least $f+1$ \echo{} messages for values other than the
most frequently occurring value
$m$ of the \echo{} messages it has received so far.
Letting $x$ be the number of \echo{} messages received for $m$,
it follows that at least $x+1$ of the \echo{} messages for values other than
$m$ are from correct processes.
But they cannot all be for the same value since no value occurs more
frequently than $m$
(cf.\ Lemma~\ref{lem:3f-binding:echo}.)

Since a correct process approves a value when it gets $n-f$ \echo{} messages
for it, the validity of the \echo{} messages implies validity of the
approved values.  In addition, the approved sets of all the correct processes
are eventually the same.
(Cf.\ Lemma~\ref{lem:3f-binding:approved}.)
The \echob{} messages preserve the validity properties of the \echo{}
messages and also ensure that the values sent in them end up being approved
(cf.\ Lemma~\ref{lem:3f-binding:echo2}).
The \echoc{} messages add a ``uniqueness'' property, ensuring at most one
non-$\bot$ value is sent by correct processes.
They also satisfy a modified approval property:  if $v$ is sent by
a correct process and $v$ is not $\bot$, then eventually every correct
process approves $v$, otherwise (if $v = \bot$) eventually every correct
process either approves $\bot$ or approves at least two values
(cf.\ Lemma~\ref{lem:3f-binding:echo3}).

We can now prove correctness and complexity when $R = 1$.
{\em Validity} is immediate from the validity properties ensured
by the \echo{}* messages.
{\em Agreement} follows from {\em Binding}
(cf.\ Proposition~\ref{prop:binding-agreement}), which we discuss next.
The first correct process to decide receives $n-f$ \echoc{} messages,
at least $n-2f$ of which are from correct processes.
If any of these messages are for a value in $V$, then by uniqueness
of \echoc{} messages, no correct process can send an \echoc{} message
for a different non-$\bot$ value, and thus no such value can be decided
subsequently.
If all of these messages are for $\bot$, then other processes can receive at
most $2f$ \echoc{} messages for any $v \in V$ ($f$ from the correct processes
that did not send \echoc{} for $\bot$ and $f$ from the faulty processes),
which is not enough support for deciding $v$ subsequently.

We next address {\em Termination and time complexity}.\footnote{
    In the full correctness proof, we first show that the algorithm terminates,
    before we scale all the message delays by the duration of the longest one.}
We first show that every correct process sends an \echob{} message by
time 2.
If at least $f+1$ correct inputs are the same, say $v$, then by time 1,
every correct process receives the initial \echo{} messages for $v$,
and sends an \echo{} message for $v$ if it has not already done so.
Thus every correct process receives at least $n-f$ \echo{} messages by time
2, and sends \echob{}.

The more interesting case, which only arises when $|V| > 2$, is when
no value occurs at least $f+1$ times among the inputs of the correct processes.
Let $x_i$ (resp., $y_i$) be the number of \echo{} messages for $v_i$ received
by a correct process $p$ from correct (resp., faulty) processes by time 1,
$1 \le i \le |V|$.
Note that $x_i \le f$ and that $x_1 + \ldots + x_{|V|} \ge n-f$ since
$p$ has received all the \echo{} messages sent by the correct processes
initially.
Let $v_m$ be the value that occurs most frequently among all the \echo{}
messages received by time 1 (breaking ties arbitrarily).
Then the total number of \echo{} messages minus the number of
those for $v_m$ is
\begin{align*}
\left( { \sum_{i=1}^{|V|} (x_i + y_i) } \right)- (x_m + y_m)
          &= \sum_{i=1}^{|V|} x_i
           + \sum_{\substack{i=1 \\ i\ne m}}^{|V|} y_i
           - x_m
      \ge (n-f) - f 
      \ge f+1 ,
\end{align*}
since $n > 3f$.
Thus $p$ sends an \echo{} message for $\bot$ by time 1
if it hasn't already done so,
and so by time 2, every correct process receives $n-f$ \echo{} messages
for $\bot$ and sends \echob{} for $\bot$ if it has not already sent an
\echob{} message.

We next show that every correct process $p$ sends an \echoc{} message by
time 4.
We rely on the fact that if $v$ is in a correct process' approved set at
time $t$, then every correct process approved set by time $t+2$
(cf.\ Lemma~\ref{lem:3f-binding:approval-delay}),
which implies that if $v$ is sent by a correct process in an \echob{}
message, then every correct process approves $v$ by time 4
(cf.\ Lemma~\ref{lem:3f-binding:echo2-approved}).
Since \echob{} messages are sent by correct processes by time 2, at least
$n-f$ arrive at $p$ by time 3.
If they are all for a common value $v$, then $p$ sends \echoc{} for $v$
by time 3.
Otherwise, $p$ waits until it has at least two approved values.
Process $p$ must have received an \echob{} for value $v_1$ from a
correct process and an \echob{} for a different value $v_2$ from
a different correct process.
Thus $p$ approves $v_1$ and $v_2$ by time 4 and sends \echoc{} by time 4.

We use a similar argument to the previous paragraph to show that every
correct process $p$ decides by time 5.
It relies on the key fact that the approval of values sent in \echoc{}
messages by the correct processes takes place by time 5
(cf.\ Lemma~\ref{lem:3f-binding:echo3-approved}).
Thus $p$ either receives at least $n-f$ \echoc{} messages for a common
value by time 5 or has approved either $\bot$ or at least two values
by time 5, and thus decides by time 5.

The {\em message complexity} is $O(|V| \cdot n^2)$ since each correct
process sends to all the processes at most one \echo{} message for
each $v \in V$, one \echob{} message, and one \echoc{} message.

We continue to prove correctness and complexity when $R = 2$.
First note that when $R = 2$, a process sends an \echod{}
message for $v$ under exactly the same circumstances that it decides
$(v,1)$ (if $v \in V$) or $(v,0)$ (if $v = \bot$) when $R = 1$.
Thus we have the
analogous properties for \echod{} messages that we had for \echoc{}
messages (validity, uniqueness, and approval)
(cf.\ Lemma~\ref{lem:3f-binding:echo4}).
These properties also carry over to \echoe{} messages
(cf.\ Lemma~\ref{lem:3f-binding:echo5}).

For {\em Agreement}, we first recall that all the non-$\bot$ values
sent in \echod{} and \echoe{} messages are the same, call it $v$.
It remains to show that if a correct process $p$ decides $(v,2)$ then
no correct process can decide $(\bot,0)$.
Since $p$ decides $(v,2)$, it receives $n-f$ \echoe{} message for $v$.
But since $n > 3f$, it's not possible for another process to receive
$n-f$ \echoe{} messages for $\bot$, which is required for a decision of
$(\bot,0)$.

{\em Validity} follows from the validity properties of \echod{} and
\echoe{} messages.

{\em Binding} holds since the binding property for $R = 1$ implies
that there is only one possible non-$\bot$ value that can be decided
in any extension after the first correct process sends its \echod{} message.

For {\em Termination and time complexity}, we argue that every correct
process decides by time 7.

Since \echod{} messages are sent for $R = 2$ exactly when decisions are
made for $R = 1$, we know that correct processes send \echod{} by time 5.
Fortunately, the approval time of 5 for values in \echoc{} messages carries
over to \echod{} messages (cf.\ Lemma~\ref{lem:3f-binding:echo4-approved}).
Thus $p$ either receives at least $n-f$ \echod{} messages for a common
value by time 6 or has approved either $\bot$ or at least two values
by time 5, and thus sends \echoe{} by time 6.

Similarly, the approval time of 5 also holds for values in \echoe{} messages
(cf.\ Lemma~\ref{lem:3f-binding:echo5-approved}).
Thus $p$ either receives at least $n-f$ \echoe{} messages for a common
value by time 7 and decides, or has approved either $\bot$ or at least two
values by time 5.
In the latter case, since $p$ receives less than $n-f$ \echoe{} messages for $\bot$,
it receives an \echoe{} message for some $w \in V$ from a correct process $q$.
In turn, $q$ received at least $n-f$ \echod{} messages for $w$, at least
$n-2f \ge f+1$ of which are from correct processes.
Since these correct processes send their \echod{} messages for $w$
by time 5, $p$ receives them by time 6.
Thus $p$ decides by time 7.

The {\em message complexity} is still $O(|V| \cdot n^2)$ since in addition
to the messages sent when $R = 1$, each process sends to all processes
one \echod{} message and one \echoe{} message.
\end{proof}

\medskip\noindent{\bf Time complexity versus round complexity:}
The upper bounds of 5 and 7 on the {\em time complexities} for
Algorithm~\ref{alg:byz-3f-binding} are tight, as shown by an
execution described in Appendix~\ref{sec:scenario}.
The execution uses $V = \{0,1\}$ and thus it is also an execution
of Algorithm~4 in~\cite{AbrahamBDY2022efficient}, implying that the
tight time complexity of the latter algorithm is also 5,
and that of Algorithm~6 in~\cite{AbrahamBDY2022efficient} is 7.
This is in contrast to the {\em round complexities} of 4 and 6
calculated in~\cite{AbrahamBDY2022efficient} for their Algorithms 4 and 6.

\begin{figure}
\includegraphics[scale=0.4]{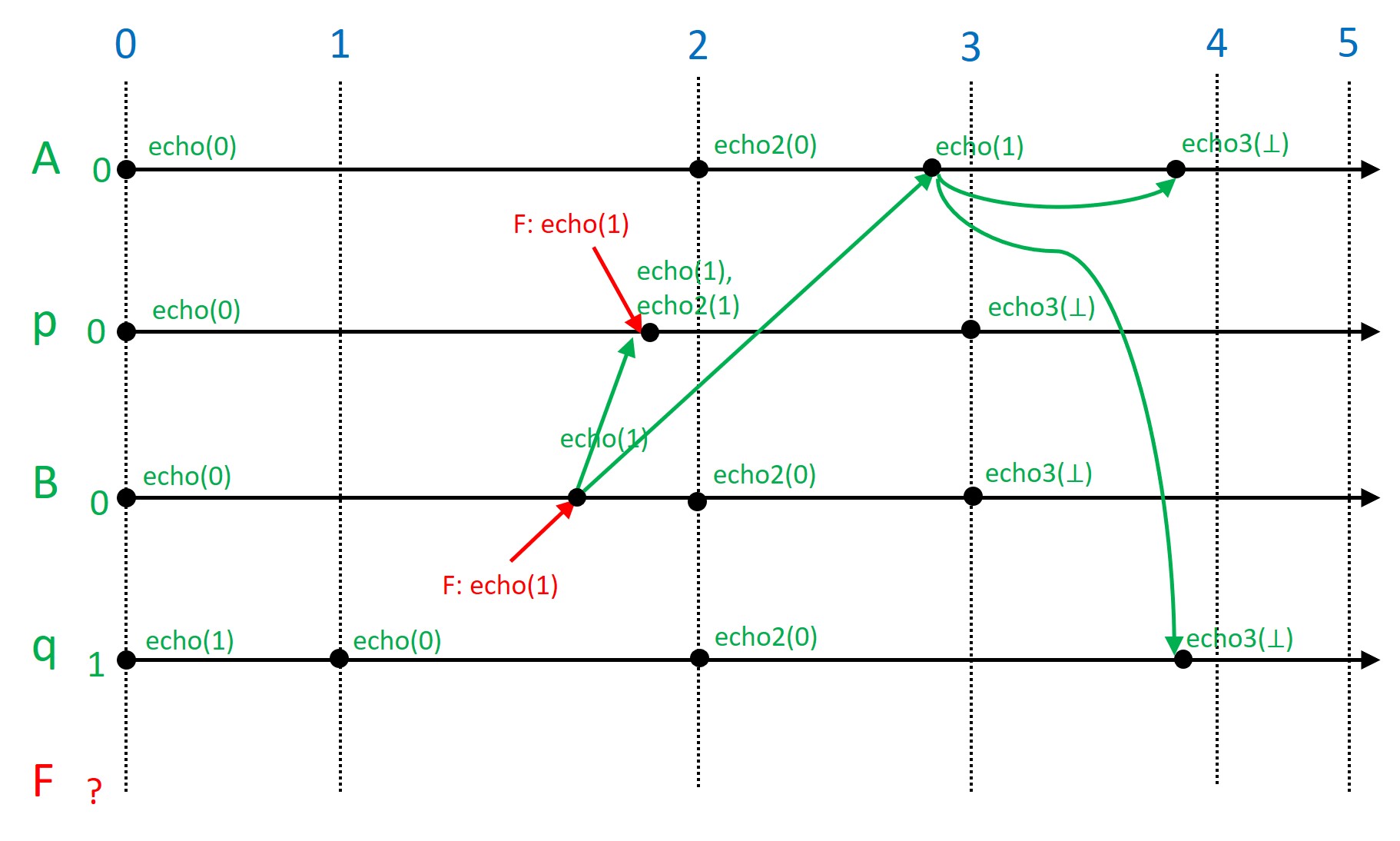}
\caption{Diagram illustrating scenario in which decision is delayed until
almost time 5.}
\label{fig:scenario-diagram}
\end{figure}

The discrepancy between round complexity and time complexity is
caused by the waiting conditions imposed before performing the next
broadcast.
If the condition is simply to receive enough messages from the previous
broadcast, then at most one time unit elapses per broadcast.
But if there is an additional condition, for instance, waiting to
approve at least two values, then the condition may take more than one
time unit to become true.
We next sketch the execution from Appendix~\ref{sec:scenario}
to illustrate this point; see also Figure~\ref{fig:scenario-diagram}.

\begin{comment}
In Appendix~\ref{sec:scenario}, we describe an execution of our
Algorithm~\ref{alg:byz-3f-binding} for $R = 1$ in which some correct
processes do not decide until arbitrarily close to time 5,
showing that our upper bound of 5 is tight.
It can be easily extended to an execution for $R = 2$
in which termination does not happen until arbitrarily close to time 7.
The execution uses $V = \{0,1\}$ and thus it is also an execution
of Algorithm~4 in~\cite{AbrahamBDY2022efficient}, implying that the
tight time complexity of the latter algorithm is also 5,
and that of Algorithm~6 in~\cite{AbrahamBDY2022efficient} is 7.
\cite{AbrahamBDY2022efficient} focuses on ``round complexity'',
which counts the number of broadcasts performed by a process; the
round complexity of their Algorithm 4 (and our $R = 1$
algorithm) is only 4; similarly, the round complexity of their
Algorithm 6 (and our $R = 2$ algorithm) is only 6.
\end{comment}

Assume 
$n = 3f+1$ and partition correct processes into sets $A$, $B$, $\{p\}$, and
$\{q\}$, where $|A| = f-1$, $|B| = f$, and let $F$ be the set of
$f$ faulty processes.
Initially all correct processes have input 0 except $q$ has input 1.
At time 0, processes send \echo{} messages with their inputs which arrive
at time 1.  At time 1, $q$ sends an \echo{} message for 0, which arrives
at time 2.
Just before time 2, every process in $B$ receives $f$ \echo{} messages for
1 from the processes in $F$, which causes it to send \echo{} for 1.
Those messages take 1 time unit to arrive at all processes except $p$,
which receives them immediately.
This then causes $p$ to send an \echo{} message for 1.
Immediately thereafter, $p$ receives $f$ \echo{} messages for 1
from the processes in $F$, which causes it to send its \echob{} message,
for 1.  These messages from $p$ take 1 time unit to arrive.
Then at time 2, all the correct processes send \echob{} message for 0,
except for $p$, which has already sent its \echob{} message (for 1).
The difficulty is that by time 3, no process has $n-f$ \echob{} messages
for a common value, due to the \echob{} message for 1 from $p$.
The processes in $B \cup \{p\}$ are not blocked from sending \echoc{}
because they have approved both 0 and 1, but the processes in $A \cup \{q\}$
have only approved 0; they are unable to approve 1 until they
get the \echo{} messages for 1 sent by the processes in $A \cup \{q\}$,
which does not happen until just before time 4.
Letting all the \echoc{} messages have delay 1 means that processes cannot
decide until shortly before time 5.

\section{Discussion}

We have proposed a new problem called connected consensus which generalizes
a number of primitives used to solve consensus, including crusader
agreement, graded broadcast, and adopt-commit, using a numeric parameter $R$.
The problem can be reduced to real-valued approximate agreement when
the input set is binary and
and to approximate agreement on graphs for multi-valued input
sets (two or more inputs).
We extended the definition of the binding property for such primitives
to the multi-valued case.

We presented efficient message-passing algorithms for connected
consensus when $R$ is~1 (corresponding to crusader agreement) or~2
(corresponding to graded broadcast), in the presence of crash and malicious
failures, for multi-valued input sets.
Our algorithms are modular in that the $R = 2$ case is obtained
by appending more communication exchanges to the $R = 1$ case.

Our algorithm for crash failures has optimal resilience and message
complexity; its time complexity is optimal for reasonable resiliencies
and improves on the best previously known algorithms, which only
handled binary inputs.

For malicious failures, we provide two algorithms that trade off
resilience against time and message complexity.
One algorithm has time complexity 1 or 2 (for $R = 1$ or $R = 2$)
and sends $O(n^2)$ messages, but requires $n > 5f$.
The other algorithm only requires $n > 3f$, but has time complexity
5 or 7 (for $R = 1$ or $R = 2$) and sends $O(|V| \cdot n^2)$ messages.
This is the same performance
as the algorithms in~\cite{AbrahamBDY2022efficient}
which are only for the case when $|V| = 2$.

The techniques used in our (simple) algorithms for crash failures
and for malicious failures with $n > 5f$
are familiar from prior work.  For example, algorithms in 
\cite{MostefaouiR2001,FriedmanMR2005} rely on similar mechanisms to solve
(standard) consensus using various kinds of oracles.
The novelty in our work is the focus on the binding property for a key 
subproblem of consensus, extracted as connected consensus.

There is a message-passing algorithm for adopt-commit
with multi-valued inputs that works in the presence of malicious failures
as long as $n > 3f$~\cite{BouzidMR2015}.
However, the number of possible inputs must be  
smaller than $\lfloor{\frac{n-(f+1)}{f}}\rfloor$,
while our algorithm works for any size input set.
The message complexity is $O(n^3)$ as compared to our $O(|V| \cdot n^2)$.
Furthermore, this algorithm avoids the challenge of ensuring the binding
property (which anyway is not well-defined for adopt-commit)
as it is combined with an oracle in order to solve consensus.

An intriguing open question is whether there is an inherent cost for
satisfying the binding property:  is there some measure, perhaps time,
in which solving connected consensus without binding is more efficient
than solving it with binding?

Adopt-commit and related primitives have been implemented in
shared-memory systems
(e.g.,~\cite{DelporteFR2021,Aspnes2012,Gafni1998,MostefaouiRRT2008})
as well as message-passing.
The connection we have made between connected consensus and approximate
agreement (on graphs) may contribute to finding improved algorithms
for these primitives in shared memory.

This connection might also be a fruitful direction for future work on
connected consensus in other timing models.
For instance, \cite{AntoniadisBDPGGVZ2023} uses adopt-commit (and variants)
to solve consensus in eventually synchronous systems.
Another interesting direction is to study connected consensus in other
fault models, such as the authenticated setting; authenticated algorithms
appear in~\cite{AbrahamBDY2022efficient} but they are for binary inputs.

\newpage
\bibliographystyle{abbrv}
\bibliography{references}

\appendix

\section{Resilience Lower Bounds}
\label{sec:resilience lb}

\noindent{\bf Lower bound of $n > 2f$ for crash failures:}
We argue that $n > 2f$ is required to solve connected consensus
with crash failures for any $R \ge 1$, even without binding.

Assume in contradiction that there is an algorithm for $n = 2f$
where $V = \{0,1\}$ and consider execution $\alpha_0$ in which all processes
have input 0 and half of the processes crash initially;
by Validity, the other half must decide $(0,R)$ by some time $t_0$.
Consider execution $\alpha_1$ in which all processes have input 1
and the other half of the processes crash initially; by Validity,
the first half must decide $(1,R)$ by some time $t_1$.
Finally consider execution $\alpha$ which is the ``merger'' of $\alpha_0$
and $\alpha_1$, in which the first half of the processes have input 0, the
other half have input 1, and messages between the halves are delayed until
after  $\max\{t_0,t_1\}$.  Since processes in the first half decide $(1,R)$
as they do in $\alpha_1$ and processes in the second half decide $(0,R)$ as
they do in $\alpha_0$, Agreement is violated.

\noindent{\bf Lower bound of $n > 3f$ for malicious failures:}
We argue that $n > 3f$ is required to solve connected consensus
for any $R \ge 1$ with malicious failures, even without binding.

To prove this lower bound, assume in contradiction there is an algorithm
that works for $n = 3f$ with $V = \{0,1\}$.
Partition the set of processes into thirds, $P_0$, $P_1$ and $P_2$,
each of size $f$.
In execution $\alpha_0$, processes in $P_0$ and $P_2$ have input 0, while
processes $P_1$ are faulty but just crash.
By some time $t_0$, processes in $P_0$ and $P_2$ decide $(0,R)$ by Validity.
In execution $\alpha_1$, processes in $P_1$ and $P_2$ have input 1, while
processes $P_0$ are faulty but just crash.
By some time $t_1$, processes in $P_1$ and $P_2$ decide $(1,R)$ by Validity.
In execution $\alpha$, processes in $P_0$ have input 0, processes in
$P_1$ have input 1, while processes in $P_2$ are malicious and act like
$P_2$ in $\alpha_0$ to $P_0$ and act like
$P_2$ in $\alpha_1$ to $P_1$, and all messages between processes in $P_0$ and
processes in $P_1$ are delayed until after $\max\{t_0,t_1\}$.
Thus processes in $P_0$ decide $(0,R)$ as they do in $\alpha_0$
and processes in $P_1$ decide $(1,R)$ as they do in $\alpha_1$, violating
Agreement.
(This lower bound can also be derived from~\cite{NowakR2019}.)

\section{Time Lower Bounds}
\label{sec:time lb}

First note that processes cannot solve connected consensus without
communicating, and thus at least one time unit is necessary.
As a result, our Algorithms~\ref{alg:crash}
and~\ref{alg:byz-5f} have optimal time complexity for $R = 1$.

In Section~\ref{subsec:time-lb-crash}, we show that two rounds are
necessary for solving connected consensus when $R = 2$ in the presence
of crash failures, as long as $n \le 4f$.
Thus our Algorithm~\ref{alg:crash} has optimal time complexity for $R = 2$
when $2f < n \le 4f$.
In contrast, when $n > 4f$, we
present a simple one-round algorithm that also satisfies Binding.

In Section~\ref{subsec:time-lb-byz}, we show that two rounds are
necessary for solving connected consensus when $R = 2$ in the presence
of Byzantine failures, as long as $n \le 9$.
Thus our Algorithm~\ref{alg:byz-5f} has optimal time complexity for $R = 2$
when $3f < n \le 9f$.
In contrast, when $n > 12f$, we
present a simple one-round algorithm; if $n > 13f$, the algorithm also
satisfies Binding.

\subsection{Crash Failures}
\label{subsec:time-lb-crash}

We use a reduction from approximate agreement and a result
in~\cite{Fekete1994asynchronous} to show that, as long as $n \le 4f$,
at least two rounds of message exchange are necessary to solve
connected consensus when $R = 2$, in the presence of crash failures.

Suppose we want to solve $\epsilon$-approximate agreement on the interval
$[0,1]$.  We show how to do so using any connected consensus algorithm
$A_{CC}$ with $V = \{0,1\}$ and
$R = \left\lceil \frac{1}{2\epsilon} \right\rceil$ that tolerates
crash failures.
Letting $v$ be the approximate agreement input for process $p$, call
$A_{CC}$ with input $v$.
To obtain the approximate agreement output from the connected consensus
output, map the $2R+1$ vertices
of the connected consensus graph, which is a chain, in order to points
in the real interval $[0,1]$ that are equally spaced, with vertex $(0,R)$
corresponding to point 0 and vertex $(1,R)$ corresponding to point 1.
Since adjacent points in $[0,1]$ are distance $\frac{1}{2R}$ apart, they are
within $\epsilon$ of each other.

For example, when $\epsilon = \frac{1}{4}$, we use connected consensus
with $R = 2$.

This transformation uses no rounds other than those used by $A_{CC}$.
Thus any lower bound on the number of rounds for approximate agreement
is also a lower bound on the number of rounds for connected consensus.

For any round-based approximate agreement algorithm,
the {\em convergence ratio} is the fraction by which the size of the interval
of values reduces after one round.
The number of rounds
needed to achieve outputs that are at most $\epsilon$ apart is
$\left\lceil \log_{\frac{1}{r}} \frac{U}{\epsilon} \right\rceil$,
where $r$ is the convergence ratio
per round and $U$ is the size of the interval of inputs.
In our case, $U = 1$ and $\epsilon = 1/4$.
Fekete proved in~\cite{Fekete1994asynchronous} that the best $r$ can be is
$\left\lceil\frac{n-f}{f}\right\rceil^{-1}$.
Thus, as long as $n \le 4f$, the number of rounds is at least 2:
\begin{align*}
\left\lceil \log_{\left\lceil \frac{n-f}{f}\right\rceil} 4 \right\rceil \ge 2
\end{align*}
implies $n \le 4f$.

Interestingly, if the resilience is higher than $4f$, a simple one-round
connected consensus algorithm with Binding is possible for $R = 2$.
The algorithm works by having processes send their inputs to each other
and wait to receive $n-f$ messages.
If there exists $v$ such that all $n-f$ messages received are for $v$, then
the process decides $(v,2)$.
Otherwise, if there exists $v$ such that at least $n-2f$ messages received are
for $v$, then the process decides $(v,1)$.
Otherwise the process decides $(\bot,0)$.
Note that since $n > 4f$, there is at most one value such that a process
receives at least $n-2f$ messages for that value.

Termination and Validity are straightforward to show.

For Agreement, first suppose that
some process $p$ decides $(v,2)$ for some $v$.  Then
$p$ receives $n-f$ messages for $v$, which means every other process
$q$ receives at least $n-2f$ messages for $v$.
Thus $q$ decides either $(v,1)$ or $(v,2)$.

Now suppose that no process decides $(v,2)$ for any $v$ but some
process $p$ decides $(v,1)$ for some $v$.  Then $p$ receives at least
$n-2f$ messages for $v$, which means every other process $q$ receives
at least $n-3f$ messages for $v$ and at most $2f$ messages for any
value $u$ other than $v$.  Since $n > 4f$, $2f$ is less than $n-2f$,
and so $q$ cannot decide $(u,1)$ for any $u$ other than $v$.  Thus $q$
decides either $(\bot,0)$ or $(v,1)$.

Finally suppose that no process decides $(v,2)$ or $(v,1)$ for any $v$.
Then every process decides $(\bot,0)$.

For the Binding property,
we again have the nice property that the locked value is
determined solely by the inputs.  In order for a process to decide
$(v,1)$ or $(v,2)$, it must receive at least $n-2f$ messages for $v$,
which means there are at least $n-2f$ inputs that are $v$.
Similarly, in order for a process
to decide $(u,1)$ or $(u,2)$ where $u$ is different from $v$, it must receive
at least $n-2f$ messages for $u$, which means there are at least $n-2f$
inputs that are $u$.
However, this would mean there are at least $2(n-2f)$ inputs.
Setting $n \ge 2(n-2f)$, we get $n \le 4f$, which is a contradiction.

\subsection{Malicious Failures}
\label{subsec:time-lb-byz}

We use the same reduction from approximate agreement as in
Section~\ref{subsec:time-lb-crash}, except that $A_{CC}$
tolerates Byzantine failures.
We appeal to a result in~\cite{DolevLPSW1986reaching} to show that, as
long as $n \le 9f$, at least two rounds of message exchange are
necessary to solve connected consensus when $R = 2$, in the presence
of malicious failures.

The relevant result in~\cite{DolevLPSW1986reaching} is that the best
the convergence ratio $r$ can be in the presence of malicious failures
is $\left\lceil \frac{n-3f}{2f}\right\rceil^{-1}$.
Plugging this value of $r$ into the formula
$\left\lceil \log_{\frac{1}{r}} \frac{U}{\epsilon} \right\rceil$,
with $U = 1$ and $\epsilon = 1/4$, results in a number of rounds that is
at least 2 as long as $n \le 9f$:
\begin{align*}
\left\lceil \log_{\left\lceil \frac{n-3f}{2f}\right\rceil} 4 \right\rceil \ge 2
\end{align*}
implies $n \le 9f$.

If the resilience is sufficiently large, $n > 12f$,
a simple one-round connected consensus algorithm is possible for $R = 2$.
Furthermore, if $n > 13f$, then the algorithm also satisfies Binding.
The algorithm is similar to that in Section~\ref{subsec:time-lb-crash}.
After receiving $n-f$ messages, each process drops the $f$ largest
and $f$ lowest values.  If all the remaining values are equal to the
same value $v$, then the process decides $(v,2)$.
Otherwise, if at least $n-6f$ of the remaining values are equal to
the same value $v$, then the process decides $(v,1)$.
Otherwise, the process decides $(\bot,0)$.

Termination is obvious.  For the rest of the correctness argument,
let $X_p$ denote the values obtained by $p$ after dropping the $f$
largest and $f$ smallest values.

\underline{Validity:}

{\em Case 1:}  $p$ decides $(v,2)$.
Then $p$ receives at least $n-3f$ messages for $v$,
at least $n-4f > 1$ are from correct processes, and thus $v$ is a
correct input.

{\em Case 2:}  $p$ decides $(v,1)$.
Then $p$ receives at least $n-6f$ messages for $v$,
at least $n-7f > 1$ are from correct processes, and thus $v$
is a correct input.
Since $p$ does not decide $(v,2)$, at least one entry in $X_p$ is not $v$.
Without loss of generality, suppose it is some $u < v$.
Thus $p$ receives at least $f+1$ messages
for values less than $v$, at least one of which is from a correct process.
Thus at least one correct input is not $v$.

{\em Case 3:}  $p$ decides $(\bot,0)$.
No value occurs in $X_p$ at least $n-6f$ times.
Let $v_{min}$ (resp., $v_{max}$) be the smallest (resp., largest) value in
$X_p$.  Since $|X_p| = n-3f$,
$v_{min} < v_{max}$ (they are different).  Thus
$p$ receives at least $f+1$ messages for
values $\le v_{min}$, and at least $f+1$ messages for values $\ge v_{max}$.
Thus there are at least two different correct inputs, one at most $v_{min}$
and one at least $v_{max}$.

\underline{Agreement:}

{\em Case 1:} Correct process $p$ decides $(v,2)$.
Then $X_p$ consists solely of $v$'s.  Thus $p$ receives
at least $n-3f$ messages for $v$, at least $n-4f$ of which
are from correct processes.
Because of the asynchrony, correct process $q$ receives at least
$n-5f$ messages for $v$ from
correct processes.  If trimming $X_p$ removes copies of $v$ from both sides,
then $X_q$ contains all $v$'s and $q$ decides $(v,2)$.
If trimming $X_p$ remoes copies of $v$ from at most one side, $X_q$
contains at least $n-6f$ $v$'s, and thus $q$ decides either $(v,1)$ or $(v,2)$.

{\em Case 2:}  Correct process $p$ decides $(v,1)$.
Then $X_p$ contains at least $n-6f$ $v$'s, meaning
$p$ receives at least $n-7f$ messages for $v$ from correct processes.  Thus
correct process $q$ receives at least $n-8f$ messages for $v$
from correct processes.
If trimming $X_p$ removes copies of $v$ from both sides, then
$X_q$ contains only $v$'s
and $q$ decides $(v,2)$.  If trimming $X_p$ removes copies of $p$ from
at most one side,
then $X_q$ contains at least $n-9f$ $v$'s.  Suppose in contradiction $X_q$
also contains at least $n-6f$ values for some $u$ other than $v$.  Then
the size of $X_q$, $n-3f$, must be at least $(n-9f) + (n-6f)$.  Algebra shows
that $n \le 12f$, a contradiction.

\underline{Binding:}
Suppose $n > 13f$.  We show that the locked value is determined solely
by the inputs.  Consider a particular input assignment.
For process $p$ to decide $(v,1)$ or $(v,2)$ in some execution with that
input assignment, it must
receive at least $n-6f$ messages for $v$, at least $n-7f$ of which are
from correct processes.
For a process $q$ to decide $(u,1)$ or $(u,2)$ where $u \ne v$ in some
execution with that input assignment, it must receive at least $n-6f$
messages for $u$, at least $n-7f$ of which are from correct processes
(different from the previous set of correct processes).  So the
number of correct processes, $n-f$, must be
at least $2(n-7f)$, which means that $n \le 13f$, a contradiction.

\section{Correctness of Algorithm~\ref{alg:crash} for Crash Failures}
\label{sec:crash proof}

This section presents the full proof of
Theorem~\ref{thm:alg-crash-correct}.

\crashtwo*

\begin{proof}
\underline{{\em Termination} holds for $R = 1$ and $R = 2$} since at most $f$
processes are faulty and messages sent by correct processes are
guaranteed to arrive.

\underline{{\em Complexity} for $R = 1$ and $R = 2$:}
Since each process sends a message to every process either once
(for $R = 1$) or twice (for $R = 2$), the message complexity is $O(n^2$).
For $R = 1$, each process decides after receiving $n-f$ round-1 messages,
which take at most 1 time unit to arrive.
For $R = 2$, each process sends its round-2 message by time 1 (after
receiving $n-f$ round-1 messages), and then decides after receiving at
most $n-f$ round-2 messages, which arrive by time 2.

\begin{lemma}
\label{lem:inputs-branch-crash}
Let $p$ be any process.
\begin{itemize}
\item[(a)] If $p$'s branch variable equals $v$ at the end of
  round 1, then at least $n-f$ processes have input $v$.
\item[(b)] If $p$'s branch variable equals $\bot$ at the end
  of round 1, then not all processes have the same input.
\item[(c)] If $p$'s branch variable equals $u \in V$ at the end of round 1
and another process $q$'s branch variable equals $v \in V$ at the end
of round 1,then $u = v$.
\end{itemize}
\end{lemma}

\begin{proof}
(a) If $p$ sets its branch variable to $v$, then all $n-f$ \inputm{}
messages it receives are for $v$.
Since processes fail only by crashing, the $n-f$ senders of these messages all
have input $v$.

(b) If $p$ sets its branch variable to $\bot$, then it receives \inputm{}
messages for more than one value.  Since processes fail only by crashing,
the values in the \inputm{} messages are the senders' inputs.

(c) If $p$ sets its branch variable to $u \in V$, then it receives
$n-f$ \inputm{} messages for $u$, and similarly $q$ receives $n-f$
\inputm{} messages for $v$.
Since $n > 2f$ and processes fail only by crashing, it must be that $u = v$.
\end{proof}

\underline{{\em Agreement for $R = 1$:}}
By Proposition~\ref{prop:binding-agreement}, this follows from Binding for
$R = 1$, which is shown below.

\underline{{\em Validity for $R = 1$:}}
If process $p$ decides $(\bot,0)$, then its branch variable at the end of
round 1 equals $\bot$.  Lemma~\ref{lem:inputs-branch-crash}(b) implies
there are at least two inputs and so vertex $(\bot,0)$ is in the minimal
subtree.
If $p$ decides $(v,1)$, then its branch variable at the end of round 1
equals $v$.  Lemma~\ref{lem:inputs-branch-crash}(a) implies that $v$
is an input and so vertex $(v,1)$ is in the minimal subtree.

\underline{{\em Binding for $R = 1$:}}
We show a stronger property, that the branch along which decisions are
made is determined solely by the inputs.
For any assignment of inputs to the processes,
since $n > 2f$, there is at most one input value $v \in V$ that occurs at least
$n-f$ times.
By Lemma~\ref{lem:inputs-branch-crash}(a), no process can set its branch
variable to, and thus decide, any value in $V$ other than $v$, in any
execution with those inputs.

\underline{{\em Agreement for $R = 2$:}}
We show that the processes' decisions label vertices in the minimal subtree
that are at distance at most 1 from each other.  In particular, we show
\begin{itemize}
\item[(i)] If a process decides $(v,1)$ or $(v,2)$,
then no process can decide $(u,1)$ or $(u,2)$ for any $u \ne v$.
\item[(ii)] If a process decides $(v,2)$ for some $v \in V$, then no
process can decide $(\bot,0)$.
\end{itemize}

(i) If process $p$ decides $(v,2)$, then all its received \branch{}
messages are for $v$ and thus all senders of these messages set their
branch variables to $v$ at the end of round 1.
If $p$ decides $(v,1)$ in Line~\ref{line:alg-crash:dec-1a},
then it receives a \branch{} message for $v$ and thus
the branch variable of the sender of that message is $v$ at the end
of round 1.
If $p$ decides $(v,1)$ in Line~\ref{line:alg-crash:dec-1b},
then $v$ is the value of its branch variable at the end of round 1.
In all three cases, Lemma~\ref{lem:inputs-branch-crash}(c) implies that
no \branch{} messages are sent for any value in $V$ other than $v$.
Thus no process can decide $(u,1)$ or $(u,2)$ for $u \ne v$.

(ii) Suppose in contradiction process $p$ decides $(v,2)$ for
some $v \in V$ and process $q$ decides $(\bot,0)$.  Then $p$
receives $n-f$ \branch{} messages for $v$ and $q$ receives $n-f$
\branch{} messages for $\bot$.
Since processes fail only by crashing, the set of processes sending
the $v$ \branch{} messages to $p$ and the set of processes sending
the $\bot$ \branch{} messages to $q$ are disjoint.
Thus $n \ge 2(n-f)$, which implies $n \le 2f$, a contradiction.

\underline{{\em Validity for $R = 2$:}}
If process $p$ decides $(\bot,0)$, then its branch variable equals $\bot$
at the end of round 1.  Lemma~\ref{lem:inputs-branch-crash}(b) implies
there are at least two inputs and so vertex $(\bot,0)$ is in the minimal
subtree.

If process $p$ decides $(v,1)$ or $(v,2)$, then $p$ receives at least
one \branch{} message for $v$.  Thus some process has its branch
variable equal to $v$ at the end of round 1.
Lemma~\ref{lem:inputs-branch-crash}(a) implies that $v$ is an input
and so leaf vertex $(v,2)$ is in the minimal subtree.
This implies Validity when $p$ decides $(v,2)$.

Suppose $p$'s decision is $(v,1)$.  To ensure Validity, we show that
there is another input besides $v$, i.e., another leaf vertex in the
minimal subtree besides $(v,2)$.
If the decision is taken in Line~\ref{line:alg-crash:dec-1a}, then $p$'s
branch variable is $\bot$ at the end of round 1 and
Lemma~\ref{lem:inputs-branch-crash}(b) implies there is another input value
besides $v$.
If the decision is taken in Line~\ref{line:alg-crash:dec-1b}, then the
\branch{} messages received by $p$ have different values.
If one of the values is some $u \in V$ that is not $v$, then
$u$ is another input value.
Otherwise one of the values must be $\bot$, which means that the sender's
branch variable equals $\bot$ at the end of round 1.
Lemma~\ref{lem:inputs-branch-crash}(b) implies there are at least
two input values.

\underline{{\em Binding for $R = 2$:}}
The Binding property for $R = 1$ implies that for every execution that
ends when the first process decides, there is at most
one $v \in V$ that can be decided in any future extension of the
execution.
For $R = 1$, processes decide on their branch variables,
implying that in any future extension, every process that sets its
branch variable sets it to $v$ or $\bot$.
For $R = 2$, processes exchange their branch variables in the second
round.  The only possible non-$\bot$ value that can be exchanged is $v$,
so the only possible decisions are $(\bot,0)$, $(v,1)$ and $(v,2)$,
which proves Binding for $R = 2$.
\end{proof}

\section{Correctness of Algorithm~\ref{alg:byz-5f} for Malicious Failures and $n > 5f$}
\label{sec:5f proof}

This section presents the full proof of Theorem~\ref{thm:alg-5f-correct}.

\malfive*

\begin{proof}
\underline{{\em Termination} holds for $R = 1$ and $R = 2$} since at most $f$
processes are faulty and messages sent by correct processes are
guaranteed to arrive.

\underline{{\em Complexity} for $R = 1$ and $R = 2$:}
Since each correct process sends a message to every process either once
(for $R = 1$) or twice (for $R = 2$), the message complexity is $O(n^2$).
For $R = 1$, each correct process decides after receiving $n-f$ round-1
messages, which take at most 1 time unit to arrive.
For $R = 2$, each correct process sends its round-2 message by time 1 (after
receiving $n-f$ round-1 messages), and then decides after receiving at
most $n-f$ round-2 messages, which arrive by time 2.

\begin{lemma}
\label{lem:inputs-branch-5f}
Let $p$ be a correct process.
\begin{itemize}
\item[(a)] If $p$'s branch variable equals $v$ at the
  end of round 1, then at least $n-4f$ correct processes have input $v$,
  at least $n-3f$ correct processes have input at most $v$, and
  at least $n-3f$ correct processes have input at least $v$.
\item[(b)] If $p$'s branch variable equals $\bot$ at
  the end of round 1, then not all correct processes have the same input.
\item[(c)] If $p$'s branch variable equals $u \in V$ at the end of round 1
  and another correct process $q$'s branch variable equals $v \in V$
  at the end of round 1, then $u= v$.
\end{itemize}
\end{lemma}

\begin{proof}
(a) If $p$ sets its branch variable to $v$, then all
$n-3f$ values remaining after dropping the $f$ smallest and $f$
largest of the $n-f$ received \inputm{} messages are equal to $v$.
Since at most $f$ of these are from faulty processes, at least
$n-4f$ correct processes have input $v$.  Ignoring the $f$ largest
values results in $n-2f$ values that are at most $v$, and at least
$n-3f$ of them are from correct processes.  Similarly, ignoring the
$f$ smallest values results in $n-2f$ values that are at least $v$,
and at least $n-3f$ of them are from correct processes.

(b) Suppose $p$ sets its branch variable to $\bot$.
Then its $W$ variable contains at least two values, say $u$ and $v$
with $u < v$.  Thus $p$ receives at least $f+1$ \inputm{} messages for
values that are at most $u$ and at least $f+1$ \inputm{} messages for
values that are at least $v$.  Thus at least one correct process has
an input that is at most $u$ and at least one correct process has an
input that is at least~$v$.

(c) Suppose in contradiction $p$ sets its branch variable to $u \in V$
and another correct process $q$ sets its branch variable to $v \in V$
with $u < v$.  Then $p$ receives at least $n-2f$ \inputm{} messages
for values that are at most $u$, while $q$ receives at least $n-2f$
\inputm{} messages for values that are at least $v$.  At least $n-3f$
of the messages received by $p$ that are at most $u$ come from correct
processes, while at least $n-3f$ of the messages received by $q$ that
are at least $v$ come from a different set of correct processes.  Thus
the total number of processes, $n$, is at least $2(n-3f)$ plus the $f$
faulty processes.  That is, $n \ge 2(n-3f) + f$, which implies $n \le
5f$, a contradiction.
\end{proof}

\underline{{\em Agreement for $R = 1$:}}
By Proposition~\ref{prop:binding-agreement}, this follows from Binding for
$R = 1$, which is shown below.

\underline{{\em Validity for $R = 1$:}}
If correct process $p$ decides $(\bot,0)$, then its branch variable at
the end of round 1 equals $\bot$.  Lemma~\ref{lem:inputs-branch-5f}(b)
implies not all correct processes have the same input and so vertex
$(\bot,0)$ is in the minimal subtree.
If $p$ decides $(v,1)$, then its branch variable at the end of round 1
equals $v$.  Since $n > 5f$, Lemma~\ref{lem:inputs-branch-5f}(a)
implies that $v$ is the input of at least one correct process and so
vertex $(v,1)$ is in the minimal subtree.

\underline{{\em Binding for $R = 1$:}}
We show a stronger property, that the branch along which decisions are
made is determined solely by the inputs.
Given an assignment of inputs to the processes,
suppose there is one execution in which a correct process decides $u \in V$
and another execution in which a correct process decides $v \in V$
with $u < v$.
By Lemma~\ref{lem:inputs-branch-5f}(a), the decision of $u$ implies
that at least $n-3f$ correct processes have inputs that are at most $u$.
Similarly, the decision of $v$ implies that at least $n-3f$ correct
processes have inputs that are at least $v$.
Thus the total number of processes $n$ is at least $2(n-3f)$ plus
the $f$ faulty processes.  That is, $n \ge 2(n-3f) + f$, which implies
$n \le 5f$, a contradiction.

\underline{{\em Agreement for $R = 2$:}}
We show that the correct processes' decisions label vertices in the
minimal subtree that are at distance at most 1 from each other.
In particular, we show
\begin{itemize}
\item[(i)] If a correct process decides $(v,1)$ or $(v,2)$,
then no correct process can decide $(u,1)$ or $(u,2)$ for any $u \ne v$.
\item[(ii)] If a correct process decides $(v,2)$ for some $v \in V$, then no
correct process can decide $(\bot,0)$.
\end{itemize}

(i) If correct process $p$ decides $(v,2)$, then it receives at least
$n-2f > 3f$ \branch{} messages for $v$, implying at least one correct
process sets its branch variable to $v$ at the end of round 1.
If $p$ decides $(v,1)$ in Line~\ref{line:alg-5f:dec-1a}, then $p$ receives
at least $f+1$ \branch{} messages for $v$ and at least one correct
process sets its branch variable to $v$ at the end of round 1.
If $p$ decides $(v,1)$ in Line~\ref{line:alg-5f:dec-1b}, then $v$ is
the value of its branch variable at the end of round 1.
In all three cases,
Lemma~\ref{lem:inputs-branch-5f}(c) implies that every correct process that
sets its branch variable to a non-$\bot$ value at the end of round 1
sets it to $v$.
Thus no correct process can receive more than $f$ \branch{} messages for
any $u \in V$ other than $v$ and thus it cannot decide $(u,1)$ or $(u,2)$.

(ii) Suppose correct process $p$ decides $(v,2)$ for some $v \in V$.
Then $p$ receives at least $n-2f$ \branch{} messages for $v$.
At least $n-3f$ of the \branch{} messages received by $p$ are from
correct processes and least $n-4f$ of these messages are also received
by every other correct process $q$.
Since $n > 5f$, $n-4f > f$ and thus $q$ receives at least $f+1$
\branch{} messages for $v$.  Thus it is impossible for $q$ to decide
$(\bot,0)$.

\underline{{\em Validity for $R = 2$:}}
If correct process $p$ decides $(\bot,0)$, then its branch variable
equals $\bot$ at the end of round 1.
Lemma~\ref{lem:inputs-branch-5f}(b) implies not all correct processes
have the same input and so vertex $(\bot,0)$ is in the minimal
subtree.

If $p$ decides $(v,2)$, then $p$ receives at least
$n-2f > 3f$ \branch{} messages for $v$.
Thus at least one correct process has its branch
variable equal to $v$ at the end of round 1.
Lemma~\ref{lem:inputs-branch-5f}(a) implies that $v$ is the input
of at least one correct process since $n > 5f$,
and so leaf vertex $(v,2)$ is in the minimal subtree.

Suppose $p$ decides $(v,1)$ in Line~\ref{line:alg-5f:dec-1a}.
Then $p$ sets its branch variable to $\bot$ at the end of round 1,
which by Lemma~\ref{lem:inputs-branch-5f}(b) implies there are at
least two correct inputs.
In addition, $p$ receives at least $f+1$ \branch{} messages for $v$,
at least one of which is from a correct process $q$.
Thus $q$ sets its branch variable to $v$ at the end of round 1,
which by Lemma~\ref{lem:inputs-branch-5f}(a) and the fact that $n > 5f$
implies that at least one correct process has input $v$.
Thus leaf vertex $(v,2)$ and at least one other leaf vertex is in
the minimal tree, implying so is vertex $(v,1)$.

Suppose $p$ decides $(v,1)$ in Line~\ref{line:alg-5f:dec-1b}.
Then $p$ sets its branch variable to some $v \ne \bot$ at the end of
round 1 but no value appears at least $n-2f$ times in its received
\branch{} messages.
By Lemma~\ref{lem:inputs-branch-5f}(a), since $n > 5f$ at least $f+1$
correct processes have input $v$ and thus leaf vertex $(v,2)$ is in
the minimal tree.
At most $n-2f-1$ of the $n-f$ \branch{} messages received by $p$
are for $v$, which leaves at least $(n-f) - (n-2f-1) = f+1$ messages
for values other than $v$.
At least one of these is from a correct process $q$ for some value $u \ne v$
which is the value of $q$'s branch variable at the end of round 1.
If $u \in V$, then Lemma~\ref{lem:inputs-branch-5f}(a) implies that some
correct process has input $u$, and thus leaf vertex $(u,2)$ is also in
the minimal tree.
If $u = \bot$, then Lemma~\ref{lem:inputs-branch-5f}(b) implies
not all correct processes have the same input and thus there is
some leaf vertex in addition to $(v,2)$ in the minimal subtree.
In all cases, vertex $(v,2)$ and another leaf vertex are in the minimal
subtree, implying so is vertex $(v,1)$.

\underline{{\em Binding for $R = 2$:}}
The Binding argument for $R = 1$ implies that for a given assignment
of inputs, there is only one possible $v \in V$ that can appear in correct
processes' branch variables at the end of round 1 in any execution.
Thus no correct process can get more than $f$ \branch{} messages for
any $u \in V$ other than $v$ and thus it cannot decide $(u,1)$ or $(u,2)$
in any execution.
\end{proof}

\section{Correctness of Algorithm~\ref{alg:byz-3f-binding} for
Malicious Failures with $n > 3f$}
\label{sec:3f proof}

This section presents the full proof of Theorem~\ref{thm:alg-3f-binding-correct}.

\malthree*

\begin{proof}
First note that because of the use of the sent\_echo variables, each
correct process sends at most one \echo{} message for each $v \in V \cup
\{\bot\}$ and at most one \echob{}, \echoc{}, and \echod{} message.
Thus we can assume some bookkeeping mechanism that drops any duplicate
messages that might come from the faulty processes.
The rest of the proof relies on the fact that a process receives at
most one of each type of message from each process.

We start with some lemmas about the \echo, \echob, and \echoc{} messages
and the approved variable.

\begin{lemma}
\label{lem:3f-binding:echo}
Let $p$ be a correct process.
\begin{itemize}
\item[(a)] If $p$ sends an \echo{} message for $v \in V$, then some correct
           process has input $v$.
\item[(b)] If $p$ sends an \echo{} message for $\bot$, then not all the
           correct processes have the same input.
\end{itemize}
\end{lemma}

\begin{proof}
(a) Suppose in contradiction $p$ sends an \echo{} message for some $v \in V$
such that no correct process has input $v$, and that $p$ is the first
correct process to do so.
The reason $p$ sends the \echo{} message is that it gets at least $f+1$
\echo{} messages for $v$.
But at least one of them must be from a correct process
contradicting the assumption that $p$ is the first correct process to
send an \echo{} message for $v$.

(b) Suppose in contradiction $p$ sends an \echo{} message for $\bot$
but all the correct processes have the same input.
Let $p$ be the first correct process to do so.
Then it cannot be that $p$ sends the \echo{} message for $\bot$
because it gets at least $f+1$ \echo{} messages for $\bot$.
Instead $p$ gets at least $f+1$ \echo{} messages for values other than the
mode\footnote{the value occurring most frequently, breaking ties
arbitrarily} $m$
of the \echo{} messages it has received so far.  Let $x =$ num\_echo$[m]$.
Then the sum of num\_echo$[w]$ for all $w \ne m$ is at least $x+f+1$.  At
least $x+1$ of the echoes for values other than $m$ are from correct
processes but they cannot all be for the same value since $x$ is the
number of times that the mode appears.  Thus not all the correct
processes have the same input.
\end{proof}

\begin{lemma}
\label{lem:3f-binding:approved}
Let $p$ be a correct process.
\begin{itemize}
\item[(a)] If $v \in V$ is an element of $p$'s approved set, then some correct
   process has input $v$.
\item[(b)] If $\bot$ is an element of $p$'s approved set, then not all
   correct processes have the same input.
\item[(c)] If $v \in V \cup \{\bot\}$ is in $p$'s approved set, then eventually
   $v$ is in the approved set of any other correct process.
\end{itemize}
\end{lemma}

\begin{proof}
(a) Suppose $v \in V$ is in $p$'s approved set.
Then $p$ receives at least $n-f$ \echo{} messages for $v$, at least $n-2f$
of which are from correct processes.
By Lemma~\ref{lem:3f-binding:echo}(a), some correct process has input $v$.

(b) Suppose $\bot$ is in $p$'s approved set.
Then $p$ receives at least $n-f$ \echo{} messages for $\bot$, at least
$n-2f$ of which are from correct processes.
By Lemma~\ref{lem:3f-binding:echo}(b), not all correct processes have
the same input.

(c) Since $v$ is in $p$'s approved set, $p$ receives at least $n-f$
\echo{} messages for $v$, at least $n-2f$ of which are from correct processes.
Thus every correct process receives at least $n-2f \ge f+1$ \echo{} messages
for $v$ and sends an \echo{} message for $v$ (if it has not already done so).
As a result, every correct process receives at least $n-f$ \echo{} messages
for $v$ and adds $v$ to its approved set.
\end{proof}

\begin{lemma}
\label{lem:3f-binding:echo2}
Let $p$ be a correct process.
\begin{itemize}
\item[(a)] If $p$ sends an \echob{} message for $v \in V$, then some correct
           process has input $v$.
\item[(b)] If $p$ sends an \echob{} message for $\bot$, then not all the
           correct processes have the same input.
\item[(c)] If $p$ sends an \echob{} message for $v \in V \cup \{\bot\}$,
           then eventually every correct process has $v$ in its approved set.
\end{itemize}
\end{lemma}

\begin{proof}
(a) Suppose $p$ sends an \echob{} message for $v \in V$.
Then $p$ receives at least $n-f$ \echo{} messages for $v$, at least
$n-2f$ of which are from correct processes.
By Lemma~\ref{lem:3f-binding:echo}(a), some correct process has input $v$.

(b) Suppose $p$ sends an \echob{} message for $\bot$.
Then $p$ receives at least $n-f$ \echo{} messages for $\bot$, at least
$n-2f$ of which are from correct processes.
By Lemma~\ref{lem:3f-binding:echo}(b),
not all the correct processes have the same input.

(c) When $p$ sends \echob{} for $v$, it has just added $v$ to its
approved set.
By Lemma~\ref{lem:3f-binding:approved}(c), $v$ is eventually in
every correct process' approved set.
\end{proof}

\begin{lemma}
\label{lem:3f-binding:echo3}
Let $p$ be a correct process.
\begin{itemize}
\item[(a)] If $p$ sends an \echoc{} message for $v \in V$, then some correct
           process has input $v$.
\item[(b)] If $p$ sends an \echoc{} message for $\bot$, then not all the
           correct processes have the same input.
\item[(c)] If $p$ sends \echoc{} for $u \in V$ and another correct process
           $q$ sends \echoc{} for $v \in V$, then $u = v$.
\item[(d)] Suppose correct process $p$ sends \echoc{} for $v$.  If $v
           \in V$, then eventually $v$ is in every correct process' approved
           set.  If $v = \bot$, then eventually either $\bot$ or at least two
           values are in every correct process' approved set.
\end{itemize}
\end{lemma}

\begin{proof}
(a) Suppose $p$ sends an \echoc{} message for $v \in V$.
Then $p$ receives at least $n-f$ \echob{} messages for $v$, at least
$n-2f$ of which are from correct processes.
By Lemma~\ref{lem:3f-binding:echo2}(a), some correct process has input $v$.

(b) Suppose $p$ sends an \echoc{} message for $\bot$.  One possibility
is that $p$ receives at least $n-f$ \echob{} messages for $\bot$, at least
$n-2f$ of which are from correct processes, in which case
Lemma~\ref{lem:3f-binding:echo2}(b) implies that not all the correct processes
have the same input.
The other possibility is that $p$'s approved set contains more than one
element, in which case Lemma~\ref{lem:3f-binding:approved}(a) and (b)
imply that not all the correct processes have the same input.

(c) Suppose $p$ sends \echoc{} for $u \in V$ and $q$ sends \echoc{}
for $v \in V$ with $u \ne v$.  Then $p$ gets at least $n-f$ \echob{}
messages for $u$, at least $n-2f$ of which are from correct processes,
and $q$ gets at least $n-f$ \echob{} messages for $v$, at least
$n-2f$ of which are from correct processes.
Since each correct process sends at most one \echob{} message,
$n \ge 2(n-2f)+f$, which implies $n \le 3f$, a contradiction.

(d) Suppose $p$ sends \echoc{} for $v$ because it receives at least $n-f$
\echob{} messages for $v \in V \cup \{\bot\}$,
at least $n-2f$ of which are from correct processes.
Then Lemma~\ref{lem:3f-binding:echo2}(c) implies eventually
$v$ is in every correct process' approved set.
Suppose $p$ sends \echoc{} for $v = \bot$ because it has more than one
approved value.
By Lemma~\ref{lem:3f-binding:approved}(c), eventually those elements
are in every correct process' approved set also.
\end{proof}

\underline{{\em Agreement for $R = 1$:}}
By Proposition~\ref{prop:binding-agreement}, this follows from Binding for
$R = 1$, which is shown below.

\underline{{\em Validity for $R = 1$:}}
Suppose $p$ decides $(v,1)$ where $v \in V$.
Then it receives at least $n-f$ \echoc{} messages for $v$, at least
$n-2f$ of which are from correct processes and
Lemma~\ref{lem:3f-binding:echo3}(a) implies that some correct process
has input $v$.
Thus vertex $(v,1)$ is in the minimal subtree.

Suppose $p$ decides $(\bot,0)$ because it receives at least $n-f$
\echoc{} messages for $\bot$.
Since at least $n-2f$ are from correct processes,
Lemma~\ref{lem:3f-binding:echo3}(b) implies that not all correct processes
have the same input.
Suppose $p$ decides $\bot$ because it has more than one approved value.
Then Lemma~\ref{lem:3f-binding:approved}(a) and (b)
imply that not all the correct processes have the same input.
In both cases, the center vertex $(\bot,0)$ is in the minimal subtree.

\underline{{\em Termination for $R = 1$:}}
We first show that eventually every correct
process sends an \echob{} message\footnote{
This is obvious when $|V| = 2$,
but here we argue that the mechanism adopted from MV-bcast~\cite{MostefaouiR2017}
ensures it even when $|V| > 2$.}.
The trigger for sending an \echob{} message is receiving at least $n-f$
\echo{} messages for a common value.
If any correct process $p$ receives at least $n-f$ \echo{}
messages for a value $v$, then every correct process $q$ receives at least
$n-f$ \echo{} messages for $v$.
The reason is that at least $n-2f$ \echo{} messages
for $v$ received by $p$ are from correct processes, so every correct process
also gets those messages.  Since $n-2f \ge f+1$, every correct process sends
an \echo{} message for $v$, and thus $q$ receives at least $n-f$ \echo{}
messages for $v$.

Suppose in contradiction no correct process $p$ ever receives at least
$n-f$ \echo{} messages for a common value.
Consider the point at which some correct process $p$ has received its last
\echo{} message and let $m$ be the mode
of the values in these messages (the value that occurs most frequently,
breaking ties arbitrarily).
Since $p$ receives at least $n-f$ \echo{} messages and at most $f$ of them
are for the same value (the mode $m$), $p$ receives at least $n-2f$ \echo{}
messages for values other than $m$.  Since $n > 3f$, it follows that
$n-2f \ge f+1$,
and thus the condition for sending an \echo{} message for $\bot$ is satisfied.
That is, every correct process sends \echo{} for $\bot$
and every correct process receives $n-f$ \echo{} messages for $\bot$,
causing it to send \echob{} for $\bot$.

Next we show that every correct process $p$ sends an \echoc{} message.
One possibility is that $p$ receives $n-f$ \echob{} messages for
a common value.
Suppose that does not occur; then we must argue that $p$ gets at least two
approved values.
Since $p$ never receives $n-f$ \echob{} messages for the same
value, at least one correct process sends \echob{} for $v_1$ and another
correct process sends \echob{} for $v_2$, $v_1 \ne v_2$.
By Lemma~\ref{lem:3f-binding:echo2}(c),
eventually $p$ has two approved values and sends \echoc{} for $\bot$.

Finally we show that every correct process $p$ decides.  One possibility
is that $p$ receives $n-f$ \echoc{} messages for a common value.  Suppose
that does not occur; then we must argue that $p$ gets at least two
approved values or approves $\bot$.
Since $p$ never receives $n-f$ \echoc{} messages for the same
value, at least one correct process sends \echoc{} for $v_1$ and
another correct process sends \echoc{} for $v_2$, $v_1 \ne v_2$.
By Lemma~\ref{lem:3f-binding:echo3}(d), eventually $p$ has two approved values
or approves $\bot$, and $p$ decides.

\underline{{\em Binding for $R = 1$:}}
Let $p$ be the first correct process to decide.  Then $p$ receives
at least $n-f$ \echoc{} messages, at least $n-2f$ of which are from correct
processes.  If at least one of the \echoc{} messages from correct processes
is for $v \in V$, then
by Lemma~\ref{lem:3f-binding:echo3}(c),
no correct process sends \echoc{} for any value in $V$ other than $v$.
Thus no process can decide $(u,1)$ for any $u \in V$ other than $v$ in
any extension of the execution ending with $p$'s decision.
Now suppose that all the \echoc{} messages received by $p$ from correct
processes are for $\bot$.
Then the maximum number of \echoc{} messages that a correct process can
receive for any value in $V$ is $2f$, $f$ from the faulty processes
and $f$ from the correct processes that did not send \echoc{} for $\bot$.
But $2f < n-f$ since $n > 3f$, and that is not enough for any correct
process to decide $(u,1)$ for any $u \in V$ in any extension of the
execution ending with $p$'s decision.

\underline{{\em Complexities for $R = 1$:}}
As remarked at the beginning of the proof, each correct process sends
at most one \echo{} message to all processes for each $v \in V \cup
\{\bot\}$ for a total of $O(|V| \cdot n^2)$, and each correct process
sends at most one \echob{} message and at most one \echoc{} message to
all processes for a total of $O(n^2)$.
The grand total is $O(|V| \cdot n^2)$.

For the time complexity, we show that every correct process decides by
time 5.

We first show that each correct process $p$ sends
its \echob{} message by time 2.

{\em Case 1:}  At least $f+1$ correct inputs are the same, call it $v$.
By time 1, every correct process $p$ receives the initial \echo{} messages
sent by the correct processes when they first wake up.
If $p$ has not already sent an \echo{} message for $v$,
then it sends one by time 1.
Thus by time 2, every correct process receives at least $n-f$ \echo{}
messages for $v$.
If $p$ has not already sent an \echob{} message, then it sends one
for $v$ by time 2.

{\em Case 2:} No value occurs more than $f$ times among the correct inputs.
Let $x_i$ (resp., $y_i$) be the number of \echo{} messages for $v_i$ received
by $p$ from correct (resp., faulty) processes by time 1, $1 \le i \le |V|$.
Note that $x_i \le f$ for all $i$ by the case assumption
and that $x_1 + \ldots + x_{|V|} \ge n-f$ since $p$ has received all the
\echo{} messages sent by the correct processes when they first wake up.
Let $v_m$ be the mode
among all the \echo{} messages received by time 1.
Then the total number of \echo{} messages minus the number of
those for $v_m$ is
\begin{align*}
\left( {\sum_{i=1}^{|V|} (x_i + y_i)} \right) - (x_m + y_m)
          &= \sum_{i=1}^{|V|} x_i
           + \sum_{\substack{i=1 \\ i\ne m}}^{|V|} y_i
           - x_m \\
      &\ge (n-f) - f = n-2f \ge f+1 \text{~~~~since $n > 3f$.}
\end{align*}
Thus $p$ sends an \echo{} message for $\bot$ if it hasn't already done so,
and so by time 2, every correct process receives $n-f$ \echo{} messages
for $\bot$ and sends \echob{} for $\bot$ if it has not already sent an
\echob{} message.

We use the following two lemmas to show that every correct process $p$
sends its \echoc{} message by time 4.
They are the timed analogs of Lemma~\ref{lem:3f-binding:approved}(c)
and Lemma~\ref{lem:3f-binding:echo2}(c).

\begin{lemma}
\label{lem:3f-binding:approval-delay}
If correct process $p$ has $v \in V \cup \{\bot\}$ in its approved
set at time $t$, then $v$ is in the approved set of every correct process
by time $t+2$.
\end{lemma}

\begin{proof}
Since $v$ is in $p$'s approved set at time $t$, $p$ has received
at least $n-f$ \echo{} messages for $v$ by time $t$.
At least $n-2f \ge f+1$ of these \echo{} messages are from correct
processes, and thus every correct process receives at least $f+1$
\echo{} messages for $v$ by time $t+1$.
Thus every correct process sends an \echo{} message for $v$ by time $t+1$,
and so every correct process receives at least $n-f$ \echo{} messages for
$v$, and adds $v$ to its approved set, by time $t+2$.
\end{proof}

\begin{lemma}
\label{lem:3f-binding:echo2-approved}
If correct process $p$ sends \echob{} for $v \in V \cup \{\bot\}$,
then every correct process has $v$ in its approved set by time 4.
\end{lemma}

\begin{proof}
We showed above that $p$ sends \echob{} by time 2.
Since $p$ sends \echob{} immediately after approving its first value,
the lemma follows from Lemma~\ref{lem:3f-binding:approval-delay}.
\end{proof}

Suppose in contradiction that correct process $p$ has not sent an
\echoc{} message by time 4.
Since every correct process sends its \echob{} message by time 2,
$p$ receives them all by time 3.
However, less than $n-f$ of them are for a common value.
Then $p$ must have received an \echob{} for some $v_1$ sent by a
correct process and an \echob{} for some $v_2$ sent by another correct
process, where $v_1 \ne v_2$.
By Lemma~\ref{lem:3f-binding:echo2-approved}, both $v_1$ and $v_2$
are in $p$'s approved set by time 4.
Thus $p$ sends its \echoc{} message (for $\bot$) by time 4.

The next lemma is used to show that every correct process decides by time 5.
It is the timed analog of Lemma~\ref{lem:3f-binding:echo3}(d).

\begin{lemma}
\label{lem:3f-binding:echo3-approved}
Suppose correct process $p$ sends \echoc{} for $v$.
If $v \in V$, then $v$ is in every correct process' approved set by time 4.
If $v = \bot$, then either $\bot$ or at least two values are in every
correct process' approved set by time 5.
\end{lemma}

\begin{proof}
Suppose $p$ sends its \echoc{} message because it receives at least $n-f$
\echob{} messages for $v$ (this is the case if $v \in V$ and might also
be the case if $v = \bot$).
Then $p$ receives at least $n-f$ \echob{} messages,
at least $n-2f$ of which are from correct processes.
By Lemma~\ref{lem:3f-binding:echo2-approved}, $v$ is in the approved
set of all correct processes by time 4.

Suppose $p$ sends its \echoc{} message (for $v = \bot$) at some time $t$
because it approves a second value.
If $t \ge 3$, then $p$ will have received all the \echob{} messages from
correct processes but there will not be $n-f$ for a common value.
Thus there are two \echob{} messages sent by correct processes for
different values.  By Lemma~\ref{lem:3f-binding:echo2-approved},
at least two different values are in the approved set of all correct
processes by time 4.
If $t < 3$, then Lemma~\ref{lem:3f-binding:approval-delay} implies
that both values in $p$'s approved set are in every process' approved set
by time at most $t+2 < 5$.
\end{proof}

Suppose in contradiction that correct process $p$ has not decided by time 5.
Since every correct process sends its \echoc{} message by time 4,
$p$ receives them all by time 5.
However, less than $n-f$ of them are for a common value.
Then $p$ must have received an
\echoc{} for some $v_1$ sent by a correct process
and an \echoc{} for some $v_2$ sent by another correct process, where
$v_1 \ne v_2$.
By Lemma~\ref{lem:3f-binding:echo3}(c), one of $v_1$ and $v_2$ is $\bot$.
Lemma~\ref{lem:3f-binding:echo3-approved} implies that either $\bot$
or at least two values are in $p$'s approved set by time 5, and
thus $p$ decides by time 5.

-----------------------------------------------

The next lemmas, concerning \echod{} and \echoe{} messages, refer to the
part of the algorithm that is only executed when $R = 2$.

\begin{lemma}
\label{lem:3f-binding:echo4}
Let $p$ be a correct process.
\begin{itemize}
\item[(a)] If $p$ sends an \echod{} message for $v \in V$,
           then some correct process has input $v$.
\item[(b)] If $p$ sends an \echod{} message for $\bot$, then not all the
           correct processes have the same input.
\item[(c)] If $p$ sends an \echod{} message for $u \in V$ and another
           correct process $q$ sends an \echod{} message for $v \in V$,
           then $u = v$.
\item[(d)] Suppose correct process $p$ sends \echod{} for $v$.  If $v
           \in V$, then eventually $v$ is in every correct process' approved
           set.  If $v = \bot$, then eventually either $\bot$ or at least two
           values are in every correct process' approved set.
\end{itemize}
\end{lemma}

\begin{proof}
Parts (a) and (b) follow from the Validity property for $R = 1$,
and part (c) follows from the Agreement property for $R = 1$,
since the value a process sends in an \echod{} message is the value
it would have decided had $R$ been 1.

(d) Suppose $p$ sends \echod{} for $v$ because it receives at least $n-f$
\echoc{} messages for $v \in V \cup \{\bot\}$,
at least $n-2f$ of which are from correct processes.
Then Lemma~\ref{lem:3f-binding:echo3}(d) implies if $v \in V$,
eventually $v$ is in every correct process' approved set,
and if $v = \bot$, eventually either $\bot$ or at least two values
are in every correct process' approved set.
Suppose $p$ sends \echod{} for $v = \bot$ because it has more than one
approved value or $\bot$ is approved.
By Lemma~\ref{lem:3f-binding:approved}(c), eventually those elements
are in every correct process' approved set also.
\end{proof}

\begin{lemma}
\label{lem:3f-binding:echo5}
Let $p$ be a correct process.
\begin{itemize}
\item[(a)] If $p$ sends an \echoe{} message for $v \in V$,
           then some correct process has input $v$.
\item[(b)] If $p$ sends an \echoe{} message for $\bot$, then not all the
           correct processes have the same input.
\item[(c)] If $p$ sends an \echoe{} message for $u \in V$ and another
           correct process $q$ sends an \echoe{} message for $v \in V$,
           then $u = v$.
\item[(d)] Suppose correct process $p$ sends \echod{} for $v$.  If $v
           \in V$, then eventually $v$ is in every correct process' approved
           set.  If $v = \bot$, then eventually either $\bot$ or at least two
           values are in every correct process' approved set.
\end{itemize}
\end{lemma}

\begin{proof}
(a) Suppose $p$ sends an \echoe{} message for $v \in V$.
Then $p$ receives at least $n-f$ \echod{} messages for $v$, at least
$n-2f$ of which are from correct processes.
By Lemma~\ref{lem:3f-binding:echo4}(a), at least one correct process
has input $v$.

(b) Suppose $p$ sends an \echoe{} message for $\bot$ because it
receives at least $n-f$ \echod{} messages for $v$.
Since at least $n-2f$ of the messages are from correct processes,
Lemma~\ref{lem:3f-binding:echo4}(b) implies that not all the correct
processes have the same input.
If $p$ sends an \echoe{} message for $\bot$ because it has more than one
approved value or $\bot$ is approved,
then Lemma~\ref{lem:3f-binding:approved}(a) and (b) imply
that not all the correct processes have the same input.

(c) Suppose $p$ sends \echoe{} for $u \in V$ and $q$ sends \echoe{}
for $v \in V$ with $u \ne v$.
Then $p$ gets at least $n-f$ \echod{} messages for $u$,
at least $n-2f$ of which are from correct processes, and $q$ gets at
least $n-f$ \echod{} messages for $v$, at least $n-2f$ of which
are from correct processes.
Since each correct process only sends one \echod{} message, the number
of processes, $n$, must be at least $2(n-2f) + f$, which implies $n
\le 3f$, a contradiction.

(d) Suppose $p$ sends \echoe{} for $v$ because it receives at least $n-f$
\echod{} messages for $v \in V \cup \{\bot\}$,
at least $n-2f$ of which are from correct processes.
Then Lemma~\ref{lem:3f-binding:echo4}(d) implies if $v \in V$,
eventually $v$ is in every correct process' approved set,
and if $v = \bot$, eventually either $\bot$ or at least two values
are in every correct process' approved set.
Suppose $p$ sends \echoe{} for $v = \bot$ because it has more than one
approved value or $\bot$ is approved.
By Lemma~\ref{lem:3f-binding:approved}(c), eventually those elements
are in every correct process' approved set also.
\end{proof}

\underline{{\em Agreement for $R = 2$:}}
By Lemmas~\ref{lem:3f-binding:echo4}(c) and~\ref{lem:3f-binding:echo5}(c),
there exists $v \in V$ such that all \echod{} and \echoe{} messages sent
by correct processes are for $v$ or $\bot$.
Consequently, every correct process either decides
$(v,2)$ (since there cannot be $n-f$ \echoe{} messages for any value
in $V$ other than $v$), or $(v,1)$ (since there cannot be $f+1$ \echod{}
messages for any value in $V$ other than $v$) or $(\bot,0)$.  It remains to
show that it is not possible for one correct process to decide $(v,2)$
and another to decide $(\bot,0)$.

Suppose in contradiction correct process $p$ decides $(v,2)$ and
correct process $q$ decides $(\bot,0)$.  Then $p$ receives at least
$n-f$ \echoe{} messages for $v$, at least $n-2f$ of which are from
correct processes, and $q$ receives at least $n-f$ \echoe{} messages
for $\bot$, at least $n-2f$ of which are from correct processes.
Since each correct process sends only one \echoe{} message,
the number of processes, $n$,
must be at least $2(n-2f) + f$, which implies $n \le 3f$, a contradiction.

\underline{{\em Validity for $R = 2$:}} Suppose correct process $p$ decides
$(\bot,0)$.  Then it receives at least $n-f$ \echoe{} messages for
$\bot$, at least $n-2f$ of which are from correct processes.
By Lemma~\ref{lem:3f-binding:echo5}(c), not all the correct processes have
the same input and the center vertex $(\bot,0)$ is in the minimal subtree.

Suppose $p$ decides $(v,2)$.  Then it receives at least $n-f$
\echoe{} messages for $v$, at least $n-2f$ of which are from correct
processes.
Lemma~\ref{lem:3f-binding:echo5}(a) implies at least one correct
process has input $v$ and vertex $(v,2)$ is in the minimal subtree.

Suppose $p$ decides $(v,1)$.  Then it receives at least $f+1$
\echod{} messages for $v$, at least one of which is from a correct
process, and so Lemma~\ref{lem:3f-binding:echo4}(a) implies at
least one correct process has input $v$.  But $p$ also has more than
one approved value or has approved $\bot$, and so
Lemma~\ref{lem:3f-binding:approved}(a) and (b) imply
not all correct processes have the same input.
Thus vertex $(v,1)$ is in the minimal subtree.

\underline{{\em Termination for $R = 2$:}}
Note that a process sends \echod{} when $R = 2$ in exactly the same
situations in which a process decides when $R = 1$.
Thus by Termination for $R = 1$, every correct process sends an \echod{}
message.

We next show that every correct process $p$ sends an \echoe{} message.
If all correct processes send \echod{} for the same value $v$, then eventually
$p$ receives them and sends \echoe{} for $v$.
Suppose that does not occur; then we must argue that $p$ gets at least
two approved values or approves $\bot$, which will cause it to send an
\echoe{} for $\bot$.
Since $p$ never receives $n-f$ \echod{} messages for the same
value, at least one correct process sends \echod{} for $v_1$ and
another correct process sends \echod{} for $v_2$, $v_1 \ne v_2$.
By Lemma~\ref{lem:3f-binding:echo4}(d), eventually $p$ has at least two
approved values or approves $\bot$, and sends \echoe{}.

Finally, we show that every correct process $p$ eventually decides.
The preceding paragraphs imply that $p$ receives at least $n-f$ \echod{}
messages and at least $n-f$ \echoe{} messages.
Suppose in contradiction that $p$ does not decide.
Then it receives less than $n-f$ \echoe{} messages
for any fixed value in $V \cup \{\bot\}$.
Consequently, there is one correct process that sends \echoe{} for $v_1$
and another correct process that sends \echoe{} for $v_2$
with $v_1 \ne v_2$.
By Lemma~\ref{lem:3f-binding:echo5}(d), eventually $p$ has more than
one approved value or approves $\bot$.
Also since $p$ does not decide, it receives less than $n-f$ \echoe{} messages
for $\bot$, and thus it must receive at least one \echoe{} message
for some $w \in V$ from a correct process $q$.
Since $q$ sends \echoe{} for $w$, it receives at least $n-f$ \echod{}
messages for $w$, at least $n-2f \ge f+1$ of which are from correct processes.
Process $p$ also receives these \echod{} messages for $w$ from the
correct processes.
Therefore the condition on Line~\ref{line:alg-3f-binding:decide-intermediate}
eventually holds, and $p$ decides.

\underline{{\em Binding for $R = 2$:}}
By the binding property for $R = 1$, there is only one possible
non-$\bot$ value that can be decided in any extension after the first
correct process sends its \echod{} message.

\underline{{\em Complexities for $R = 2$:}}
In addition to the $O(|V| \cdot n^2)$ messages sent for the $R = 1$ portion
of the algorithm, each process sends at most one \echod{} message and
at most one \echoe{} message to all processes
as remarked at the beginning of the proof,
for a grand total of $O(|V| \cdot n^2)$.

For the time complexity, we show that every correct process decides
by time 7.
By the code, a process sends its \echod{} message under exactly the same
circumstances that it decides in the $R = 1$ case.
Thus the time complexity argument for $R = 1$ implies that each correct
process sends its \echod{} message by time 5.

We use the next lemma to show that each correct process sends its \echoe{}
message by time 6.
It is the timed analog of Lemma~\ref{lem:3f-binding:echo4}(d).

\begin{lemma}
\label{lem:3f-binding:echo4-approved}
Suppose correct process $p$ sends an \echod{} message for $v$.
If $v \in V$, then $v$ is in every correct process' approved set by time 4.
If $v = \bot$, then either $\bot$ or at least two values are in every
correct process' approved set by time 5.
\end{lemma}

\begin{proof}
Suppose $p$ sends \echod{} because it receives at least $n-f$ \echoc{}
messages for $v$, at least $n-2f$ of which are from correct processes.
Then Lemma~\ref{lem:3f-binding:echo3-approved} gives the result.

Suppose $p$ sends \echod{} (for $\bot$) because it does not have at least
$n-f$ \echoc{} messages for a common value.
Then two correct processes send \echoc{} for two different values
and again Lemma~\ref{lem:3f-binding:echo3-approved} gives the result.
\end{proof}

Suppose in contradiction that correct process $p$ has not sent an
\echoe{} message by time 6.
Since each correct process sends its \echod{} message by time 5,
$p$ has received them all by time 6.
However, less than $n-f$ of them are for a common value.
Then $p$ must have received an
\echod{} for some $v_1$ sent by a correct process
and an \echod{} for some $v_2$ sent by another correct process, where
$v_1 \ne v_2$.
By Lemma~\ref{lem:3f-binding:echo4-approved}, either $\bot$ or at
least two values are in $p$'s approved set by time 5 and thus $p$
sends \echoe{} by time 6.

The next lemma helps to show that every correct process decides by time 7.
It is the timed analog of Lemma~\ref{lem:3f-binding:echo5}(d).

\begin{lemma}
\label{lem:3f-binding:echo5-approved}
Suppose correct process $p$ sends an \echoe{} message for $v$.
If $v \in V$, then $v$ is in every correct process' approved set by time 4.
If $v = \bot$, then either $\bot$ or at least two values are in every
correct process' approved set by time 5.
\end{lemma}

\begin{proof}
Suppose $p$ sends \echoe{} because it receives at least $n-f$ \echod{}
messages for $v$, at least $n-2f$ of which are from correct processes.
Then Lemma~\ref{lem:3f-binding:echo4-approved} gives the result.

Suppose $p$ sends \echoe{} (for $\bot$) because it does not have at least
$n-f$ \echoc{} messages for a common value.
Then two correct processes send \echod{} for two different values
and again Lemma~\ref{lem:3f-binding:echo4-approved} gives the result.
\end{proof}

Suppose in contradiction that correct process $p$ has not decided by time 7.
Since every correct process sends its \echoe{} message by time 6,
$p$ receives them all by time 7.
Thus the first conjunct of the ``if'' statement on
Line~\ref{line:alg-3f-binding:decide-intermediate} holds by time 7.

We now show that the second conjunct of the ``if'' statement on
Line~\ref{line:alg-3f-binding:decide-intermediate} holds by time 7.
Since less than $n-f$ of \echoe{} messages are for a common value,
$p$ must have received an
\echoe{} for some $v_1$ sent by a correct process
and an \echoe{} for some $v_2$ sent by another correct process, where
$v_1 \ne v_2$.
By Lemma \ref{lem:3f-binding:echo5-approved}, $p$ has at least two
values or $\bot$ in its approved set by time 5.

We finish by showing that the third conjunct of the ``if'' statement in
Line~\ref{line:alg-3f-binding:decide-intermediate} holds by time 7.
Since $p$ receives less than $n-f$ \echoe{} messages
for $\bot$, it must receive at least one \echoe{} message
for some $w \in V$ from a correct process $q$.
Since $q$ sends \echoe{} for $w$, it receives at least $n-f$ \echod{}
messages for $w$, at least $n-2f \ge f+1$ of which are from correct processes.
Since these correct processes send their \echod{} messages for
$w$ by time 6, $p$ receives them by time 7.

Thus all three conjuncts of the ``if'' statement on
Line~\ref{line:alg-3f-binding:decide-intermediate} hold by time 7,
and $p$ decides by time 7.
\end{proof}

\section{Detailed Scenario with Time Complexity 5 for Algorithm~\ref{alg:byz-3f-binding}}
\label{sec:scenario}

We describe an execution of Algorithm~\ref{alg:byz-3f-binding} with $R = 1$
and $V = \{0,1\}$ in which some \echoc{} messages are sent arbitrarily
close to time 4, which delays the decisions of correct processes until
arbitrarily close to time 5.
It can be easily extended to an execution for $R = 2$
in which termination does not happen until arbitrarily close to time 7.
The execution uses $V = \{0,1\}$ and thus it is also an execution
of Algorithm~4 in~\cite{AbrahamBDY2022efficient}, implying that the
tight time complexity of the latter algorithm is also 5,
and that of Algorithm~6 in~\cite{AbrahamBDY2022efficient} is 7.

Consider any $f \ge 2$ and $n = 3f+1$.
The threshold for forwarding \echo{} messages is $f+1$
and the threshold for sending \echob{} messages and approving values is $2f+1$.
Since $f \ge 2$, $f+2$ is less than $2f+1$, which is important
for the scenario to work out.

Partition the $3f+1$ processes into:
\begin{itemize}
\item set $A$ of $f-1$ correct processes,
\item correct process $p$,
\item set $B$ of $f$ correct processes,
\item correct process $q$, and
\item set $F$ of $f$ faulty processes.
\end{itemize}

Assume that all correct processes have input 0 except $q$ which has
input 1.

At time 0, all correct processes send \echo{} messages for their inputs,
which arrive at time~1.

At time 1, each correct process has received $2f$ \echo(0)'s and 1 \echo(1).
Since $A \cup \{p\} \cup B$ already sent \echo(0) they send nothing, while $q$
sends \echo(0) which arrives at time~2.

Let $\epsilon$ be an arbitrarily small positive real number.
At time $2 - \epsilon$, the $f$ processes
in $B$ receive \echo(1) from the $f$ processes in $F$.  They now have $f+1$
\echo(1)'s so they send \echo(1).  All but one of these messages take
1 time unit to arrive (so they don't arrive until time $3 - \epsilon$;
we will return to this later).  However, the messages to $p$ arrive very
quickly, at time $2 - \epsilon/2$ (still just before time 2), causing $p$
to have $f+1$ \echo(1) messages ($f$ from $B$, 1 from $q$) and to send
\echo(1), which takes $1 - \epsilon/2$ time unit to arrive, arriving at
time $3-\epsilon$.

At time $2 - \epsilon/4$ (still before time 2), $p$ receives
\echo(1) from the $f$ processes in $F$.  So $p$ now has received $2f+1$
\echo(1) messages ($f$ from $B$, $f$ from $F$, and 1 from $q$).
So $p$ {\bf approves} 1 and sends \echob(1), which takes 1 time unit to arrive
(arriving at time $3-\epsilon/4$).

At time 2, $q$'s \echo(0) message arrives everywhere, causing all
the correct processes to have $2f+1$ echo(0)'s.  So they all {\bf approve} 0
and all send \echob(0), except for $p$, which already sent \echob(1).  These
\echob{} messages take 1 time unit, so they arrive at time 3.

At time $3-\epsilon$, the \echo(1) messages sent by $B$ arrive at all
the correct processes (except $p$, which already received them) and
the \echo(1) message sent by $p$ arrives at all the correct processes.
As a result, every process in $A$ has $f+2$ \echo(1) messages (1 from
$q$, $f$ from $B$, and 1 from $p$).  This causes processes in $A$ to
send \echo(1), but it is not large enough to cause $A$ to approve 1,
since $f+2 < 2f+1$.  These messages take 1 time unit, arriving at time
$4 - \epsilon$.

Process $p$ now has $2f+2$ \echo(1)'s, since it received its own.

Processes in $B$ now have $2f+2$ \echo(1)'s (1 from $q$, $f$ from $F$,
$f$ from $B$, and 1 from $p$).
They don't send \echob(1) because they already sent
\echob(0) but they {\bf approve} 1.

Process $q$ now has $f+2$ \echo(1)'s (1 from itself, $f$ from $B$,
and 1 from $p$).
It doesn't (re)echo 1 because it sent \echo(1) initially.  It doesn't
approve 1 because $f+2 < 2f+1$.

At time $3 - \epsilon/4$, $p$'s \echob(1) message arrives at all the
correct processes and at time 3, all the other correct processes' \echob(0)
messages arrive at all the correct processes.
So at time 3, each correct process has $2f$ \echob(0) messages and 1
\echob(1) message.
One rule for sending \echoc{} is to have $2f+1$ \echob{}
messages for a common value, which isn't the case here.  The other
rule for sending \echoc{} is to have at least two approved values.
Processes in $B \cup \{p\}$ have approved both 0 and 1, so they send
\echoc$(\bot)$.  But processes in $A \cup \{q\}$ have only approved 0, as
they have only $f+2 < 2f+1$ \echo(1)'s.

At time $4 - \epsilon$,  the \echo(1)'s sent by $A$ at time $3 - \epsilon$
arrive at all the processes.
Now the processes in
in $A \cup \{q\}$ have $2f+2$ \echo(1)'s, and they {\bf approve} 1
and send \echoc$(\bot)$ which take 1 time unit, arriving at $5 - \epsilon$.

At time 4, all the \echoc$(\bot)$ messages from $B \cup \{p\}$ arrive
at every process.  This is only $f+1$ messages, so processes cannot
decide until they get the $f+1$ \echoc$(\bot)$ messages from
$A \cup \{q\}$ at time $5 - \epsilon$.

\section{No Built-in Binding, for $n < 5f$}
\label{subsec:lb-rb}

In this section we show that the locked value for
the Binding property cannot be predetermined from the correct processes'
inputs if $n \le 5f$, even if faulty processes cannot equivocate.
Our approach is to consider any algorithm that works by having processes
obtain $n-f$ values, at most one from each process, such that
if two processes $p$ and $q$ both get values corresponding to process $r$,
then the values are the same and if $r$ is correct then that value
is $r$'s input.
Then the algorithm must decide based only on the multiset of values it
has received.
We call such an algorithm {\em uniform}.

Consider a uniform
algorithm with $n \le 5f$.
We show that if a process receives all but $f$ 0's, then it must
decide 0 and if it receives all but $f$ 1's, then it must decide 1.
Then we show that if a process receives about half 0's and half 1's,
then it must decide $(\bot,0)$ (this relies on the fact that two
configurations with different non-$\bot$ values must have more than
$f$ values that are different).  Finally we rely on the fact that the
number of values a process receives is at most $4f$ to show that if
the first process to decide gets about half 0's and half 1's and
decides $(\bot,0)$, then in one extension we can replace $f$ of the 0's
with 1's to get a configuration that requires a decision of $(1,1)$, and in
another extension we can replace $f$ of the 1's with 0's to get a
configuration that requires a decision of $(0,1)$, which violates Binding.

\begin{theorem}
\label{thm:round-lb}
If a uniform
connected consensus algorithm for $R = 1$ with
$n$ processes, up to $f$ of which can be malicious, satifies the
Binding property, then $n > 5f$.
\end{theorem}

\begin{proof}
Consider for contradiction such an algorithm with $n \le 5f$.
Without loss of generality, assume $V = \{0,1\}$.
For simplicity, we refer to the possible decisions as $0$, $1$, and $\bot$,
instead of $(0,1)$, $(1,1)$, and $(\bot,0)$.
Let $D(z) \in \{0,1,\bot\}$ denote the decision made when the
multiset of $n-f$ values received has $z$ 0's.

We first show that if there is an ``overwhelming'' number of values
received for 0, then the decision must be 0, and similarly for 1.
The threshold is $n-2f$, which is at least $f+1$ since the resilience
lower bound discussed in Appendix~\ref{sec:resilience lb} 
shows that $n$ must be at least $3f+1$.

\begin{lemma}
\label{lem:required-dec}
Let $z$ be any integer in $\{0,\ldots,n-2f\}$.
\begin{itemize}
\item[(a)] If $z \ge n-2f$, then $D(z) = 0$.
\item[(b)] If $z \le f$, then $D(z) = 1$.
\end{itemize}
\end{lemma}

\begin{proof}
(a) Consider any execution in which correct process $p$ receives
$z \ge n-2f$ 0's and $n-f-z$ 1's.  This execution is indistinguishable
from one in which all $n-f$ of the correct processes have input 0,
and $p$ receives $z \ge n-2f$ messages for 0 from the correct
processes and $n-f-z \le f$ messages for 1 from faulty processes.
By the Validity condition, $p$ must decide 0 in the second
execution.  Thus $D(z) = 0$.

(b) Consider any execution in which correct process $p$ receives
$z \le f$ 0's and $n-f-z$ 1's.  This execution is indistinguishable from
one in which all $n-f$ of the correct processes have input 1, and $p$
receives $n-f-z$ messages
for 1 from the correct processes and $z \le f$ messages
for 0 from faulty processes.  By the Validity condition,
$p$ must decide 1 in the second execution.  Thus $D(z) = 1$.
\end{proof}

\begin{lemma}
\label{lem:3f+2}
The number of processes $n$ must be at least $3f+2$.
\end{lemma}

\begin{proof}
Suppose in contradiction that $n = 3f+1$ and consider any correct
process $p$.  If the majority of the $n-f = 2f+1$ values received by
$p$ is 0, then $z \ge f+1$.  Since $n = 3f+1$, $f+1 = n-2f$, and thus
Lemma~\ref{lem:required-dec}(a) implies that $p$ must decide 0.  If
the majority value is not 0, then the majority value must be 1, and
Lemma~\ref{lem:required-dec}(b) implies that $p$ must decide 1.  Thus
there is no possibility of a process deciding $\bot$, and hence the
algorithm actually solves consensus, which is impossible~\cite{FischerLP1985}
\end{proof}

The next lemma shows that the range of input values requiring a decision of 0
and the range of input values requiring a decision of 1 must be sufficiently
separated from each other.

\begin{lemma}
\label{lem:min-diff}
Let $x$ and $y$ be integers in $\{0, \ldots, n-2f\}$.
If $D(x) = 0$ and $D(y) = 1$, then $|x-y| > f$.
\end{lemma}

\begin{proof}
Suppose in contradiction there exist $x$ and $y$ in $\{0,\ldots,n-2f\}$
such that $D(x) = 0$, $D(y) = 1$, and $|x-y| \le f$.

Without loss of generality, suppose $x > y$.  Consider the execution
in which the correct inputs are $x$ 0's and $n-f-x$ 1's.  Suppose that
correct process $p$ hears from all the correct processes, so it gets
$x$ 0's and $n-f-x$ 1's, and decides 0.  Suppose that another correct
process $q$ hears from only $y$ of the correct processes with input 0,
all $n-f-x$ of the correct processes with input 1, and $x-y \le f$
faulty processes, who pretend to have input 1.  Thus $q$ gets $y$ 0's
and $(n-f-x)+(x-y) = n-f-y$ 1's, and decides 1.  This is possible
because of the asynchrony of the message deliveries.  But this
violates the Agreement property.
\end{proof}

The next claim states that when the received values are about half 0's
and half 1's, the decision must be $\bot$.
It also shows that this situation is not that far from a situation
requiring a decision of 0 and also not that far from a situation
requiring a decision of 1.

\begin{claim}
\label{claim:mid-bot} Let $m = \left \lceil \frac{n-f}{2} \right \rceil$.
\begin{itemize}
\item[(a)] $m+f \ge n-2f$,
\item[(b)] $m-f \le f$, and
\item[(c)] $D(m) = \bot$.
\end{itemize}
\end{claim}

We show that Binding is not guaranteed.

Consider an execution $\alpha$ in which
$m = \left \lceil \frac{n-f}{2} \right \rceil$ of the correct processes
have input 0 and the remaining $n-f-m$ correct processes have input 1.  Let
correct process $p$ be the first to receive $n-f$ messages, which are
all from the correct processes.  So $p$ gets $m$ 0's and $n-f-m$ 1's.
By Claim~\ref{claim:mid-bot}(c), it decides $\bot$.

Also suppose that all messages to another correct process $q$ are delayed
until after $p$ decides and that the faulty processes do nothing until
after $p$ decides.

Let $\alpha_0$ be an extension of $\alpha$ in which $q$ receives $m$ 0's from
correct processes, $n-2f-m$ 1's from correct processes, and $f$ 0's from faulty
processes.  So $q$ receives $m+f$ 0's and $n-2f-m$ 1's.
Since $m+f \ge n-2f$ by Claim~\ref{claim:mid-bot}(a),
Lemma~\ref{lem:required-dec}(a) implies that $q$ decides 0.

Let $\alpha_1$ be an extension of $\alpha$ in which $q$ receives $m-f$
0's from correct processes, $n-f-m$ 1's from correct processes, and
$f$ 1's from faulty processes.  So $q$ receives $m-f$ 0's and $n-m$
1's.  Since $m-f \le f$ by Claim~\ref{claim:mid-bot}(b),
Lemma~\ref{lem:required-dec}(b) implies that $q$ decides 1.

The existence of extensions $\alpha_1$ and $\alpha_0$ violates the
Binding property.
\end{proof}

\end{document}